\documentclass[11pt]{article}

\usepackage{hyperref}
\usepackage{amsmath, amsfonts, amsthm, amssymb, graphicx}
\usepackage{xspace}
\usepackage{epsfig}
\usepackage{psfrag}
\numberwithin{equation}{section}

\let\a=\alpha \let\b=\beta \let\g=\gamma  \let\e=\epsilon
    
\let\l=\lambda \let\m=\mu  \let\x=\xi 
\let\s=\sigma   

    \let\G=\Gamma
\let\del=\partial
\let\cdel=\nabla
\newcommand{\hf}{\frac{1}{2}}
\newcommand{\qt}{\frac{1}{4}}

\def\dalemb#1#2{{\vbox{\hrule height .#2pt
        \hbox{\vrule width.#2pt height#1pt \kern#1pt
                \vrule width.#2pt}
        \hrule height.#2pt}}}

\def\0{{\sst{(0)}}}
\def\1{{\sst{(2)}}}
\def\2{{\sst{(2)}}}
\def\3{{\sst{(3)}}}
\def\4{{\sst{(4)}}}
\def\5{{\sst{(5)}}}
\def\6{{\sst{(6)}}}
\def\7{{\sst{(7)}}}
\def\8{{\sst{(8)}}}

\def\ep{\epsilon}
\def\td{\tilde}

\def\nn{\nonumber}

\let\pa=\partial 
\newcommand{\be}{\begin{equation}}
\newcommand{\ee}{\end{equation}}
\def\ba{\begin{array}}
\def\ea{\end{array}}

\def\del{\partial}
\def\sst#1{{\scriptscriptstyle #1}}

\def\half{{\textstyle{\frac{1}{2}}}}

\def\bo1{ \left | B^0 (p^+) \right \rangle}

\def\<{ \langle }
\def\>{ \rangle }

\def\S{\Sigma}

\newcommand{\bea}{\begin{eqnarray}}
\newcommand{\eea}{\end{eqnarray}}
\newcommand{\ra}{\rightarrow}

\newcommand{\Lra}{\Leftrightarrow}

\newcommand{\tr}{{\rm tr} }

\newcommand{\ams}
{{\it Institute for Theoretical Physics, \\
Science Park 904, Postbus 94485,
1090 GL Amsterdam, The Netherlands}
\\ { \tt R.N.Caldeira-Costa@uva.nl, M.Taylor@uva.nl}}

\newcommand{\auth}{\large R.~N.~Caldeira Costa and Marika Taylor}

\begin{document}

\begin{flushright}
\end{flushright}

\vspace{25pt}

\begin{center}

{\Large \bf  Holography for chiral scale-invariant models}

\vspace{20pt}

\auth

\vspace{15pt}


{\ams}

\vspace{20pt}

\begin{abstract}

Deformation of any $d$-dimensional conformal field theory by a constant null source for a vector operator of
dimension $(d + z -1)$ is exactly marginal with respect to anisotropic scale invariance, of dynamical exponent $z$.
The holographic duals to such deformations are AdS plane waves, with $z=2$ being the
Schr\"{o}dinger geometry. In this paper we explore holography for such chiral scale-invariant models. The special case of $z=0$
can be realized with gravity coupled to a scalar, and is of particular interest
since it is related to a Lifshitz theory with dynamical exponent two upon dimensional reduction.
We show however that the corresponding reduction of the dual field theory is along a null circle, and thus the Lifshitz
theory arises upon discrete light cone quantization of an anisotropic scale invariant field theory.

\end{abstract}

\end{center}
\newpage
\tableofcontents
\newpage

\section{Introduction}

Gauge/gravity dualities have become an important new tool in extracting strong coupling physics.
The best understood examples of such dualities involve relativistic quantum field theories.
Strongly coupled non-relativistic QFTs are common place in condensed matter physics and as
such there would be many interesting applications had one had under control holographic
dualities involving non-relativistic QFTs.
Motivated by such applications \cite{Son:2008ye,Balasubramanian:2008dm} initiated a discussion of holography for
$(d+1)$ dimensional spacetimes with metric\footnote{An earlier approach to the geometric realization of non-relativistic
symmetries can be found in \cite{Duval:1990hj} and the connection between this approach and 
holographic realizations is discussed in \cite{Duval:2008jg}.},
\be
\label{eq:nrmetric}
ds^2 = \frac{\sigma^2 du^2}{r^{2z}} + \frac{2 du dv + dx^i dx^i  + dr^2}{r^2}\,,
\ee
with $i \in \{1, \ldots, d-2\}$. The isometries of this metric form include
\begin{align}
&\mathcal H : u \rightarrow u + a, \nn \\
&\mathcal M : v \rightarrow v + a, \label{group} \\
&\mathcal D : r \rightarrow (1 - a) r, \qquad
 u \rightarrow (1-a)^{z} u, \qquad v \rightarrow (1-a)^{2-z} v, \qquad x^i \rightarrow (1-a) x^i \nn
\end{align}
along with rotations, translations and Galilean boosts in the $x^i$ directions. Here $\mathcal D$ is the generator
of non-relativistic scale transformations with dynamical exponent $z$. In the case of $z=2$ the isometry
group becomes the Schr\"{o}dinger group, which includes the additional special conformal symmetry
\be
\mathcal C : r \rightarrow (1 - a u) r, \qquad
u \rightarrow (1 - a u) u, \qquad v \rightarrow v + \frac{a}{2} (x^i x^i + r^2), \qquad x^i \rightarrow (1 - a u) x^i
\ee
Much of the interest in such holographic models has centered around this case of $z=2$, following the initial
suggestion that the metric \eqref{eq:nrmetric} could play the role of a background for the holographic study
of critical non-relativistic systems with $z=2$ in $(d-1)$ spacetime dimensions, for example fermions at unitarity, which have the same symmetry group.

The spacetime \eqref{eq:nrmetric} solves the equations of motion for gravity coupled to a massive vector field for
all $z > 0$. Working in the limit where $\sigma^2$ is small and treated as a perturbation around $AdS$, the
standard AdS/CFT dictionary shows that the dual field theory is a deformation of the conformal field theory by a vector operator. More
specifically, the dual conformal field theory is deformed by a constant null source for a vector operator ${\cal V}_v$ of scaling dimension $(d + z -1)$
\be
S_{\rm cft} \rightarrow S_{\rm cft} + \int du d v d^{d-2} x b {\cal V}_{v}. \label{def}
\ee
With respect to the relativistic scaling dimension, this deformation is relevant for $z < 1$, marginal for $z =1$ and irrelevant
for $z>1$. However, the deformation is exactly marginal with respect to the non-relativistic scaling symmetry for any $z$ and in
this paper we will explore holographic duality for these models. In the context of two dimensional conformal field theories, such
deformations have been previously considered by Cardy \cite{Cardy:1992tq} and the resulting models were called chiral scale-invariant models, a terminology
which we will adopt here\footnote{Whilst such theories are often called non-relativistic, or ${\rm Schr}(z)$, this terminology is arguably somewhat misleading;
the theory only becomes non-relativistic after compactification on a null direction.}.

In the case of $z=2$ the original goal was to model holographically a dual non-relativistic $(d-1)$ dimensional theory, in a background with
coordinates $(u,x^i)$ where $u$ plays the role of time. In this setup one considers operators ${\cal O}_{\Delta_s,m} (u,x^i)$ of definite scaling
dimension $\Delta_s$ and of charge $m$ under the symmetry ${\mathcal M}$. This charge $m$, which corresponds to momentum in the $v$ direction, would
then have to be identified with a discrete quantum number such as particle number. In order to discretize the possible values of $m$ one
therefore needs to compactify the $v$ direction in the holographic realization. This procedure is however very nontrivial as in general quantum corrections become
important and one cannot trust the metric \eqref{eq:nrmetric} with a compact null direction, see the discussions in \cite{Maldacena:2008wh}. (The problems 
in compactifying any field theory along a null direction are discussed in, for example, \cite{Hellerman:1997yu} and would in particular apply
to the field theories considered here.) Recent work aiming at
obtaining Schr\"{o}dinger solutions without such a compact direction can be found in \cite{Balasubramanian:2010uw}. For general $z$ and $\sigma^2 > 0$
one can reduce along $u$ (for $z < 1$) and $v$ (for $z > 1$) to obtain a $(d-1)$-dimensional theory with non-relativistic scale invariance; in all
cases the reduction is however null from the perspective of the dual quantum field theory.

For every value of $z$ compactification of a null direction will be associated with problems at the quantum level and
in this paper we will consider \eqref{eq:nrmetric} with both $u$ and $v$ non-compact. The effects of such a
compactification may be considered afterwards but this issue will
be for the most part suppressed. If the coordinates $(u,v)$ are non-compact, holography relates the bulk spacetime to a
$d$-dimensional theory which is not Lorentz invariant but which admits
scaling symmetry. Theories of this anisotropic scale-invariant type can certainly model interesting physical systems and have
appeared previously in the condensed matter literature. For example,
the $Z_N$ chiral Potts models were introduced to model systems with melting transitions \cite{Ostlund:1981zz,Howes:1983mk}. The isotropic $Z_N$ models
admit a continuum limit at criticality which is described by two-dimensional $Z_N$ conformal field theories \cite{Fateev:1985mm}.
Since the chiral $Z_N$ models are inherently anisotropic in their critical properties \cite{Albertini:1988ux,Baxter:1988me,Baxter:1988xk}, they
cannot be described by a conformal field theory in the continuum limit. Instead, as was shown in \cite{Cardy:1992tq} for superintegrable chiral Potts models,
their continuum limits correspond to deformations of conformal field theories of the type \eqref{def}, which are anisotropic but respect scale invariance.

The case studied in \cite{Cardy:1992tq} was the deformation of a specific two-dimensional conformal field theory by a vector operator of dimension $9/5$, which
corresponds to scale invariance with $z= 4/5$. As we show in section \ref{se:field-th}, anisotropic scale invariance constrains two point functions of scalar
operators at zero temperature to be of the form
\be
\< {\cal O}_{\Delta_{\cal D}} (k_u,k_v) {\cal O}_{\Delta_{\cal D}'} (-k_u,-k_v) \>
= k_u^{(\Delta_{\cal D} + \Delta_{\cal D}')/z} f ( b k_{\chi}), \label{gfr-mom}
\ee
where $(k_u,k_v)$ are the lightcone momenta, $\Delta_{\cal D}$ is the anisotropic scaling dimension and
$f(b k_{\chi})$ is an arbitrary function of the quantity
\be
k_{\chi} = 2^{z/2} k_v^{z/2} k_{u}^{z/2-1},
\ee
which is invariant under anisotropic scale transformations.
In \cite{Cardy:1992tq} two point functions in the deformed theory were computed to leading order in
$b$ using conformal perturbation theory; this amounts to computing the function $f(b k_{\chi})$ to first order
in the expansion in powers of $(b k_{\chi})$.

Conformal perturbation theory is restricted to weak chirality, namely since $b$ must be small, the theory must be close to
the isotropic point. Since the deformation is exactly marginal with respect to anisotropic scaling, the chirality
$b$ can be arbitrarily large, and the holographic realizations allow correlation functions to be computed in a strongly coupled theory,
at finite chirality. Quantities computed from the holographic models have certain universal features, as is typical for holography. For example,
only certain functions $f(b k_{\chi})$ are realized in these models and the ratio of $\eta/s$ for black holes in these models is the expected
$1/4 \pi$, since the background solves relativistic two derivative equations of motion.

There are several other motivations for exploring these anisotropic backgrounds. The case of $z=0$, which cannot be realized with massive vectors but can
be realized by coupling gravity to a scalar field,
is related to Lifshitz with dynamical exponent $Z_L = 2$ upon dimensional reduction. Embedding Lifshitz into string compactifications had proved elusive,
but this kind of realization can be obtained in Sasaki-Einstein reductions \cite{Donos:2010tu}.
Note that the $z=0$ anisotropic geometry is asymptotically $AdS$, but the dimensionally
reduced theory has Lifshitz symmetry; since holography for the former is well-understood, a holographic dictionary for the latter
can be obtained straightforwardly by dimensional reduction. However, as we will discuss here, the dimensional reduction is
on a null circle, and this DLCQ reduction introduces subtleties.

Another reason for studying general $z$ is the following. The case of Schr\"{o}dinger ($z=2$) has been extensively studied in previous literature, but
the encoding of  the dual stress energy tensor in the asymptotics of the bulk geometry remains elusive. As shown in \cite{Guica:2010sw} there are several
reasons for this subtlety. Firstly,
the natural operator in the anisotropic dual theory couples not to the metric, but to the vielbein. Secondly, linearized sources for the dual stress
energy tensor and
deforming vector operator blow up near the boundary of the spacetime faster than the Schr\"{o}dinger background. In \cite{Guica:2010sw} the general
linearized solution of
the metric and vector equations of motion about the Schr\"{o}dinger background was presented; this solution consisted of certain
independent `T' and `X' modes, which should
relate to the stress energy tensor and deforming vector operator respectively. In this paper we will show how
these `T' and `X' modes are related to the dual operators for
 $z < 1$ (when the spacetime is asymptotically locally anti-de Sitter) and explain what this implies for the holographic dictionary of Schr\"{o}dinger.
More generally, for $z > 2$, we demonstrate that the irrelevant nature of the deforming operator in the original CFT is reflected in the counterterm
structure of the deformed theory: an infinite series of counterterms are required to compute correlation functions in the deformed theory.

The plan of this paper is as follows. In section \ref{se:model} we introduce the massive vector models used to engineer the anisotropic geometries,
and discuss how they may be embedded into string theory. We also consider the special case of $z=0$ which can be realized using gravity coupled
to a scalar field. In section \ref{se:field-th} the field theoretic description of these models is described, and the form of the correlation functions
in the anisotropic theory is described. In section \ref{sec:hr} holographic renormalization is carried out in the case of $d=2$, resulting in a precise
map between the asymptotic geometry and boundary data. In section \ref{se:linear} two point functions of the stress energy tensor and of the deforming vector
operator in the scale invariant background are computed. In section \ref{se:blackhole} black hole solutions which are asymptotic to the anisotropic
scale invariant background are explored. In section \ref{se:conclusions} we give conclusions.

\section{Massive vector model} \label{se:model}

Consider the Lagrangian:
\be
S = \frac{1}{2 \kappa_{d+1}^2} \int d^{d+1} x \sqrt{-g} \left [ R + \Lambda
- \frac{1}{4} F_{mn} F^{mn} - \frac{1}{2} m^2 B_{m} B^{m} \right ],
\ee
where $F_{mn} = 2 \pa_{[m} B_{n]}$,
$\Lambda = d (d - 1)$ and $m^2 = z (z + d - 2)$. The field equations are
\bea
R_{mn} &=& -  d g_{mn} - \frac{1}{4 (d-1)} F^2 g_{mn} + \frac{1}{2} F_{mp}
F_{n}^{\; p} + \frac{1}{2} m^2 B_{m} B_{n}; \nn \\
D_{m} F^{mn} &=& m^2 B^{n},
\eea
where in addition $D_{m} B^{m} = 0$.

These equations of motion admit both an $AdS_{d+1}$ solution,
\be
ds^2 = \frac{d \rho^2}{4 \rho^2} + \frac{1}{\rho} \eta_{ab} (x) dx^{a} dx^{b},
\ee
in which $B_{m} = 0$ and a solution with anisotropic scale invariance:
\bea
ds^2 &=& \frac{d \rho^2}{4 \rho^2} + \frac{1}{\rho} \left ( \sigma^2 \rho^{1 -z} (du)^2
 + 2 du dv + dx^{i} dx_{i} \right ); \nn \\
B_{u} &=& b \rho^{-z/2}, \label{non-rel}
\eea
where
\be
b^2 = \frac{2 \sigma^2 (1-z)}{z}.
\ee
This solution is a special case of an AdS pp-wave solution
and it becomes $AdS_{d+1}$ when the parameter $\sigma$ is zero whilst any finite $\sigma$ can be rescaled to
one via the rescalings $u \rightarrow \sigma^{-1} u$, $v \rightarrow \sigma v$.
In addition to the rotations, translations and Galilean boosts in the $(d-2)$ spatial directions $x^i$, the isometry group of this background is:
\bea
M &:& v \rightarrow v + a, \qquad H: u \rightarrow u + a, \\
D &:& \rho \rightarrow (1- a)^2 \rho, \qquad x^i \rightarrow (1-a) x^i, \qquad v \rightarrow (1- a)^{2-z} v, \qquad u \rightarrow (1 -a)^z u. \nn
\eea
Here $D$ is the non-relativistic scaling (dilatation) symmetry. For general $z$ these are the only symmetries, but
at $z=2$ the metric admits the Schr\"{o}dinger symmetry group, which includes in addition a special conformal symmetry.

In the case of $z=1$ the vector field vanishes. The metric
\be
ds^2 = \frac{d \rho^2}{4 \rho^2} + \frac{1}{\rho} \left ( \sigma^2 (du)^2
 + 2 du dv + dx^{i} dx_{i} \right )
\ee
solves the Einstein equations with negative cosmological constant for any constant value of $\sigma^2$. Here $\sigma^2$ acts as a constant
source for the $T_{vv}$ component of the stress energy tensor. If this source is zero, the metric is pure $AdS_{d+1}$ whilst if $\sigma^2$ is
non-zero the metric admits only non-relativistic scale invariance, as the rotational symmetry is broken.

The case of $z=0$ is also special: the vector field is massless, dual to a conserved current, and adding a source for this current given by
\be
B_{u} = b,
\ee
gives no contribution to the bulk stress energy tensor, so $AdS_{d+1}$ with this vector field solves the bulk field equations
for any value of $b$. For $z=0$ and $d=2$
the constant coefficient $\sigma^2$ can however be switched on arbitrarily,
independently of $b$, and relates to the expectation value of the stress energy tensor. One can realize
$z=0$ in general dimensions by coupling gravity to a scalar field, as we will discuss below.

\bigskip

It is interesting to note that the solutions (\ref{non-rel}) also arise in topologically massive gravity (TMG) in three dimensions. The action for TMG
is
\be
S = \frac{1}{16 \pi G_N} \int d^3 x \, \sqrt{-g} \Big( R- 2\Lambda+ \frac{1}{2\m} \e^{l m n } \big( \G_{l s}^r \del_m \G_{r n}^s
+ \frac{2}{3} \G_{l s}^r \G_{m t}^{s} \G_{n r}^t \big) \Big)
\ee
where $\Gamma_{m n}^ l$ are the connection coefficients associated to the metric $g_{m n}$ and where we use the covariant $\e$-symbol
such that $\sqrt{-g}\epsilon^{m n r} = 1$, with $r$ the radial direction in \eqref{eq:nrmetric}. Variation of the action results in the equations of motion:
\be
\label{eq:eomtmg}
{R_{m n} - \hf g_{m n}R + \Lambda G_{m n} + \frac{1}{2 \m} \Big(\epsilon_m^{\phantom{m} r s}\cdel_r R_{s n}
+ \epsilon_n^{\phantom{n}rs} \cdel_r R_{s m}\Big) = 0.}
\ee
Spacetimes \eqref{non-rel} with generic $z$ can be realized as solutions of TMG: the spacetime
solves the TMG field equations when $\mu = (2 z - 1)$. These TMG solutions were discussed in \cite{Olmez:2005by} and fit into
the classification given in \cite{Chow:2009km} as pp-waves. Solutions of this type with $u$ compactified were recently discussed
in \cite{Anninos:2010pm}.

In \cite{Skenderis:2009nt,Skenderis:2009kd} details of the holographic dictionary for TMG were presented, and this dictionary reflects the various
problems of the theory: the theory is non-unitary and contains negative norm states. The most important feature of the dictionary for our purposes
is that, since the equations of motion of TMG are third order in derivatives, we need to specify not only the boundary metric but also (a component of) the
extrinsic curvature in order to find a unique bulk solution. When we apply gauge/gravity duality to TMG with a negative
cosmological constant, the extra boundary data corresponds to the source of an extra operator. Therefore, besides the boundary
energy-momentum tensor $T_{ij}$, which couples to the boundary metric $g_{(0)ij}$, we also have a new operator $X_{vv}$ which couples
to the leading coefficient of the radial expansion of the $(uu)$ component of the extrinsic curvature. It was shown in
\cite{Skenderis:2009nt} that this operator $X_{vv}$ has weights $(h_L, h_R)  = \hf (\m + 3, \m -1)$.

In order to realize the scale invariant background with exponent $z$ we need to work at $\mu = (2 z -1)$ and switch on a constant source for the operator $X_{vv}$.
However for $z < 1$, the case of primary interest in this paper, the deforming operator $X_{vv}$ has negative scaling weights in the conformal field theory.
This pathology is related to the lack of unitarity of the dual theory, and in this paper we work instead with the massive vector models which do not exhibit
such problems.

\subsection{Linearized equations of motion about AdS background}

Let us first linearize the equations of motion about the AdS background by letting $g_{ab} = \eta_{ab} + h_{ab}$; we fix radial axial
gauge for the metric fluctuations. The linearized Einstein equations
decouple from the vector field equations, and
the linearized vector field equations are solved by
\be
B_{a} = B_{(-z) a} (x^c) \rho^{-z/2}  + B_{(2-z) a} (x^c) \rho^{1-z/2} + \cdots + B_{(z+d-2)a} (x^c) \rho^{z/2+d/2 -1} + \cdots
\ee
where $(B_{(-z) a},B_{(z+d-2)a})$ are arbitrary $d$-dimensional 1-forms, and the other coefficients in the expansion are determined
in terms of these functions. The radial component of the vector field is completely determined in terms of these coefficients via
the divergence equation (\ref{div}) and the vector field equations. Note that the relation between mass and CFT operator dimension $\Delta_v$
for a vector is
\be
m^2 = (\Delta_v - 1) (\Delta_v + 1 - d),
\ee
which implies that
\be
\Delta_v = (z + d - 1)
\ee
for the operator dual to the vector field. This relation means in particular that the vector operator is irrelevant for $z > 1$ and
relevant for $z < 1$. When $z=0$, the vector field becomes massless and is dual to a conserved current. When $z=1$ the vector operator
is marginal with respect to the relativistic scaling symmetry.

Consider now the non-relativistic background (\ref{non-rel}). Suppose the parameter $b$ is small and one retains only
terms linear in $b$, so the metric is purely $AdS_{d+1}$. The linearized AdS/CFT dictionary
then implies that there is a constant null source for the dual vector operator of dimension $\Delta_v$. When the latter is irrelevant,
deforming the theory in this way changes the UV structure. The corresponding holographic statement is that at finite $b$ the spacetime ceases
to be asymptotically $AdS_{d+1}$; its asymptotic structure is modified and holography is extremely subtle. However, when the deforming
operator is relevant the spacetime remains asymptotically $AdS_{d+1}$ and the standard AdS/CFT dictionary can be developed. It is this
latter case that we will mostly focus on here, although we will extend our results to $z > 1$ wherever possible.

\subsection{Global structure of the spacetime for $z < 1$}

In this section we will briefly describe the global structure of the spacetime for $z<1$, which is analogous to that of the corresponding
spacetimes with $z > 1$. Since we are only interested in the case where $b \neq 0$,
it is convenient to absorb the parameter $b$ in the rescaling $u \to \s^{-1} u$, $v \to \s v$, and also
change the radial coordinate to $\rho = r^2$. The background metric and the vector field are then
\begin{eqnarray}
ds^{2} &=& g_{mn} dx^{m}dx^{n} = {1 \over r^2} \left( dr^{2} + 2 dudv + r^{2(1-z)} du^{2} + dx^i dx_i \right); \nn \\
B &=& {b \over \s}\, r^{-z}\, du.
\end{eqnarray}
In order to infer geodesic incompleteness, it is useful to consider the equation for null geodesics, which
satisfy the equation
\be
\dot{r}^{2} + 2\, \dot{v}\, \dot{u} + 2 \dot{x}^i \dot {x}_i + r^{2(1-z)}\, \dot{v}^{2} = 0.
\ee
This equation can be written in terms of the
constants of motion associated with the Killing vectors $k_{u} = \partial_{u}$, $k_{v} = \partial_{v}$ and $k_{i} = \partial_{x}^i$:
\begin{equation}
P_u = k_{u}^{a} \dot{x}_{a} = {\dot{v} \over r^2}; \qquad
P_v = - k_{v}^{a} \dot{x}_{a} = {\dot{u} \over r^2} + {\dot{v} \over r^{2z}}; \qquad P_i = \frac{ \dot{x^i}}{r^2},
\end{equation}
resulting in
\begin{equation}
\int_{r_{0}}^{r(\l)}
\frac{dr} {r \sqrt{P_u^{2}\, r^{2(2-z)} + 2\, r^{2}\, P_u P_v + r^2 P_i P^i} } = \pm \int_{\l_{0}}^{\l} d\tau.
\end{equation}
Provided that $2 P_{u} P_{v} + P_i P_i> 0$, null geodesics reach $r = \infty$ in real, finite affine parameter and hence the spacetime
is geodesically incomplete. However, this geodesic incompleteness will not prevent us from computing correlation functions unambiguously 
in this background, as we will see later; the situation is analogous to that found in Lifshitz spacetimes \cite{Kachru:2008yh}. Moreover, in section 6 we will
see that the geometries can be blackened, with a horizon cloaking the geodesic incompleteness.  
A singularity is considered acceptable according to the commonly applied holographic criteria discussed in 
\cite{Gubser:2000nd} if correlation functions can be computed unambiguously and the singularity can be cloaked by a horizon. 
Precisely this criterion was used in \cite{Kachru:2008yh} to argue that holography for Lifshitz spacetimes made sense, despite
the geodesic incompleteness. Applying the same criteria here, 
one can sensibly discuss holography for these spacetimes but it would of course still be desirable to understand the resolution
of this singularity at the quantum level, for example, by embedding these geometries into string theory. 

Next let us consider whether there is a well-defined time function in the spacetime. Reinstating the parameter $b$ the metric
is
\be
ds^2 = \frac{d \rho^2}{4 \rho^2} + \frac{1}{\rho} \left ( \frac{z}{2 (1-z)} b^2 \rho^{1-z} du^2 + 2 du dv + dx^i dx_i  \right)
\ee
where $b^2 > 0$ in the massive vector model. Thus $g_{uu} > 0$ (for finite $\rho$) for $z < 1$ and $g_{uu} < 0$ (for finite $\rho$) for $z > 1$, but
note that for all $z$ hypersurfaces of constant $u$ are null. In the case of $z > 1$, the $u$ coordinate has been treated as a time coordinate and 
real-time physics
has been defined with respect to this coordinate \cite{Leigh:2009eb,Barnes:2010ev}.
However, the fact hypersurfaces of constant $u$ are actually null is symptomatic of a larger issue:
there is no global time function in these spacetimes and the spacetimes are said to be causally non-distinguishing, which in turn implies
subtleties in treating modes of zero lightcone momentum \cite{Blau:2010fh}.

In the case where $z < 1$, one might similarly suppose that the
$u$ coordinate should be treated as spacelike, but note that hypersurfaces of constant $u$ are still null and $u$ is a null coordinate
in the background for the dual quantum field theory. Unlike the $z > 1$ case, there is a global time function: the spacetime
is asymptotically anti-de Sitter, and the coordinate defined as
\be
t = \frac{1}{\sqrt{2}} (v - u)
\ee
is everywhere timelike for $b^2 > 0$ and $z < 1$, since hypersurfaces of constant $t$ are everywhere spacelike.
Unlike the case of $z > 1$, there are no subtleties in addressing real-time physics, and
real-time issues will be suppressed here.

\subsection{Embedding of massive vector models into string theory}

One may next wonder whether these massive vector models can be realized in string theory compactifications. In the case of $z>1$, various embeddings
into string theory have been found, with the massive vector actions arising as consistent truncations of type II supergravities, see for example
\cite{Maldacena:2008wh,Hartnoll:2008rs,Gauntlett:2009zw,Donos:2009en,Colgain:2009wm,Bobev:2009mw,
Donos:2009xc,Donos:2009zf,Singh1,Singh2}. Note that in these cases the truncation to a graviton and massive vector suffices for the zero temperature background, but
additional scalar fields are switched on in the corresponding black hole solutions. From the consistent truncation perspective, it is only consistent
for the scalar fields to vanish when the vector field is null.

A necessary condition for an embedding of $0 < z  < 1$ into string theory to exist would be that there is a vector of
mass squared $0 < m^2 < (d-1)$ (in AdS units) in the spectrum around an $AdS_{d+1}$ solution, corresponding to a vector operator
in the dual CFT$_d$ of dimension $(d-1) < \Delta < d$. However, in spherical compactifications, the dimensions of all vector operators
dual to supergravity modes are necessarily integral; this follows from
the eigenvalue spectra of operators on the sphere, see for example \cite{Kim:1985ez} for the $S^5$ compactification of type IIB.
Whilst spherical compactifications includes vectors dual to symmetry currents of dimension $(d-1)$ and can
include vectors of dimension $d$ also, the chiral spectrum does not include non-integral dimension vectors.

Irrational values for the conformal dimensions of operators dual to supergravity modes in Sasaki-Einstein compactifications are however generic. As
an example, let us consider the $T^{1,1}$ compactification of type IIB supergravity, whose spectrum was computed in \cite{Ceresole:1999zs,Ceresole:1999ht}.
Since $T^{1,1}$ is a rank one $SU(2)^2/U(1)$ coset,
all differential operators can be expressed in terms of the Laplace-Beltrami operator, which is the only functionally
independent differential operator. This property allows one to compute the complete KK spectrum in this case, whilst for generic Sasaki-Einstein
compactifications only a subset of the KK spectrum is known. All masses are expressible in terms of the scalar Laplacian eigenvalue:
\be
H_{0}(j,l,r) = 6 [ j (j+1) + l (l+1) - \frac{1}{8} r^2 ],
\ee
where $(j,l,r)$ refer to the $SU(2)^2$ and R-symmetry quantum numbers. The supergravity compactification consists of graviton multiplet, four gravitino
multiplets and four vector multiplets, for which the conformal dimensions of the dual operators are expressible in terms of the function $H_0$. These conformal
dimensions are generically irrational. In particular, considering one of the vector multiplets, the corresponding dual operator to the vector is of
dimension
\be
\Delta = -1 + \sqrt{4 + H_{0}(j,l,r)}.
\ee
In special cases where the square root assumes a rational value the dual operator will have a rational conformal dimension, and will form part of a shortened
multiplet. Generically, however, the dimension will be irrational and the operator will be part of a massive long multiplet. For the chiral model to be realized,
we would need the spectrum to contain a vector operator of dimension $3 < \Delta < 4$. From \cite{Ceresole:1999zs,Ceresole:1999ht}, one finds that
vector operators with protected dimensions do not realize any operators with dimension $3 < \Delta < 4$, although both $\Delta =3$ and $\Delta = 4$ do occur.
This is in agreement with the fact that $0 < z < 1$ solutions were not found in the systematic explorations of \cite{Donos:2009xc,Donos:2009zf}.
However, since for general compactifications there is no supersymmetry or unitarity
obstruction to such operators being contained in the spectrum, it would be interesting to explore
further whether embeddings of these models into such string compactifications exist.

\subsection{Realization of $z=0$ with scalar fields} \label{z0}

In general dimensions, the case of $z=0$ realized with gravity coupled to a gauge field is special, since the gauge field corresponding to
a constant null source for the dual current does not backreact on the metric. However, $z=0$ can also be realized by coupling
gravity and a cosmological constant to a massless scalar field; as we will now discuss, this case is related to the supergravity
solutions found recently in \cite{Donos:2010tu}.

Consider first the Lagrangian:
\be
S = \frac{1}{2 \kappa_{d+1}^2} \int d^{d+1} x \sqrt{-g} \left [ R + \Lambda
- \frac{1}{2} (\pa \Phi)^2 \right ],
\ee
where $\Lambda = d (d - 1)$. The field equations are
\be
R_{mn} = -  d g_{mn} + \pa_{m} \Phi \pa_{n} \Phi; \qquad \Box \Phi = 0. \label{up}
\ee
As well as the $AdS_{d+1}$ solution with constant scalar field they also admit a solution with non-relativistic scale invariance $z=0$:
\bea
ds^2 &=& \frac{d \rho^2}{4 \rho^2} + \frac{1}{\rho} \left ( \sigma^2 \rho (dF)^2
 + 2 du dv + dx^{i} dx_{i} \right ); \nn \\
\Phi &=& \sqrt{(d-2)} \sigma F(u), \label{non-rel-z0}
\eea
where $F(u)$ is an arbitrary function of $u$\footnote{After this paper was published, we became aware of related discussions of
null scalar deformations in the earlier work \cite{Balasubramanian:2010uk}. See also the more recent work \cite{Narayan:2011az}.} 
The scalar field vanishes in $d=2$, where an arbitrary value of $\sigma^2$ satisfies
the Einstein equations with negative cosmological constant. In this case $\sigma^2$ corresponds to a non-vanishing expectation value
of $T_{vv}$, and the geometry is dual to a specific state in the conformal field theory, rather than to a
non-relativistic deformation of the original conformal field theory.

A massless field $\Phi$ is dual to a marginal scalar operator ${\cal O}_{\Phi}$ in the conformal field theory. A non-vanishing $\sigma$
implies that there is a $u$-dependent source for the dual operator, so the deformed theory is:
\be
S_{\rm CFT} \rightarrow S_{\rm CFT} + \sqrt{(d-2)} \sigma \int du dv d^{d-2} x F(u) {\cal O}_{\Phi}.
\ee
A priori it is not obvious that such deformations are exactly marginal with respect to the $z=0$ scaling symmetry. When the function $f(u)$ is constant,
the deformation does not break Lorentz symmetry, but the marginal scalar operator is not generically exactly marginal
with respect to the relativistic scaling symmetry. For general $f(u)$ the deformation respects $z=0$ symmetry under which
$v \rightarrow \lambda^2 v$, $u \rightarrow u$, $x^i \rightarrow \lambda x^i$, given that the scalar operator has non-relativistic scaling
dimension equal to the relativistic scaling dimension $d$. One would however still need to show that the scaling dimension remains exactly marginal
under the deformation and hence that the deformed theory remains scale invariant; this proof will be discussed in the next section.

This system is particularly interesting for the following reason. If one considers the case where $dF = du$, the metric can be written as
\be
ds^2 = \frac{d \rho^2}{4 \rho^2} + \frac{1}{\rho} \left ( dx_i dx^i - \sigma^{-2} \frac{dv^2}{\rho} \right ) + \sigma^2 (du + \rho^{-1} \sigma^{-2} dv)^2.
\label{red1}
\ee
Dimensionally reducing along the $u$ direction results in a $d$-dimensional metric with vector field $A$,
\bea
ds_{d}^2 &=& \frac{d\rho^2}{4 \rho^2} +  \frac{1}{\rho} \left ( dx_i dx^i - \sigma^{-2} \frac{dv^2}{\rho} \right ); \\
A &=& \frac{dv}{\sigma^2 \rho},
\eea
which exhibits Lifshitz symmetry with dynamical exponent $Z_L=2$ and corresponds to the massive vector model used to obtain Lifshitz solutions
in \cite{Taylor:2008tg}. The Lifshitz symmetry group with dynamical exponent $Z_L$ includes a dilatation symmetry
\be
\rho \rightarrow \lambda^2 \rho; \qquad
v \rightarrow \lambda^{Z_L} v; \qquad
x^i \rightarrow \lambda x^i,
\ee
and $v$ is a time coordinate. Note however that strictly speaking the scalar field in (\ref{non-rel-z0}) cannot
be dimensionally reduced along the $u$ direction, as $F(u) = u$. Whilst the $d$-dimensional vector and metric, together with
the constraint that $d \Phi = \sqrt{d-2} \sigma$, are sufficient to solve the $(d+1)$-dimensional equations of motion, it would be desirable
to find an explicit realization of a $z=0$ system in string theory and, if possible, a consistent truncation to $(d+1)$-dimensional equations of
motion.

Such families of solutions were found in Sasaki-Einstein compactifications in \cite{Donos:2010tu}. In particular, compactifications of type IIB
on Sasaki-Einstein manifolds $E_5$ admit solutions in which the ten-dimensional metric is
\be
ds^2 =  \frac{d \rho^2}{4 \rho^2} + \frac{1}{\rho} \left ( f \rho (du)^2
 + 2 du dv + dx^{i} dx_{i} \right ) + ds^2 (E_5)
\ee
where $f$ is in general a function of both Sasaki-Einstein coordinates and of $u$. The corresponding five-form $F_5$, the complex three-form $G$
and the complex one-form $P$ are respectively
\bea
F_5 &=& du \wedge dv \wedge d (\rho^2) \wedge dx_1 \wedge dx_2 + 4 {\rm Vol}_{E)5}; \\
G_3 &=& du \wedge W; \qquad P = g d \sigma, \nn
\eea
where $W$ and $g$ are a three-form and a function defined on $E_5$ which may also depend on $u$. The equations of motion imply that
\bea
d u \wedge d W &=& d_{\ast_E} W = 0; \\
- \Box_{E_5} f + 4 f &=& 4 |g|^2 + |W|^2. \nn
\eea
In general the function $f$ depends both on $u$ and on the Sasaki-Einstein coordinates.
There is a simpler subclass of solutions in which $f$ is constant and the metric becomes the product of (\ref{red1}) with a Sasaki-Einstein space.
We can furthermore consider the case where the axion and dilaton is trivial, and so $g=0$
In this case the solutions require that
\be
4 f = |W|^2,
\ee
with $W$ a harmonic form on the Sasaki-Einstein. To make contact with the discussion above it is useful to let $f = \sigma^2$, so that
\bea
ds^2 &=& \frac{d \rho^2}{4 \rho^2} + \frac{1}{\rho} \left ( \sigma^2 \rho (du)^2
 + 2 du dv + dx^{i} dx_{i} \right ) + ds^2 (E_5); \\
F_5 &=& du \wedge dv \wedge d (\rho^2) \wedge dx_1 \wedge dx_2 + 4 {\rm Vol}_{E)5}; \nn \\
G_3 &=& 2 \sigma du \wedge \td{W}, \nn
\eea
where $W = 2 \sigma \td{W}$ and hence
$|\td{W}|^2 = 1$. In the limit where $\sigma$ is small, we may analyze the interpretation of the solution using the standard AdS/CFT
dictionary. From the form of $G_3$ one can see that at order $\sigma$
it indeed corresponds to switching on a $u$ dependent source for a dimension four scalar operator in the dual four-dimensional CFT. Moreover,
suppose one considers the reduction
\bea
ds^2 &=& ds^2 (M_5)  + ds^2 (E_5); \\
F_5 &=& 4 ( {\rm Vol} (M_5) + {\rm Vol} (E_5)); \nn \\
H &=& \sqrt{2} d \Phi \wedge \td{W}. \nn
\eea
The equations of motion for
the metric on the five-dimensional non-compact manifold $M_5$ and the scalar $\Phi$ are precisely those given in \eqref{up}, but
in order to satisfy the ten-dimensional equations of motion, one needs to impose the additional constraint
\be
\partial^m \Phi \partial_{m} \Phi = 0,
\ee
and thus the reduction is not technically a consistent reduction. A similar issue was found in \cite{Donos:2010tu} in
reducing the system further to four dimensions, retaining only the four-dimensional metric and massive vector.
A consistent truncation to four dimensions involving additional fields was presented in
\cite{Donos:2010tu}.

To summarize, for the cases described in \cite{Donos:2010tu} where the function $f$ is independent of the Sasaki-Einstein,
the corresponding $(d-1)$-dimensional holographic theory should be the dimensional reduction along the $u$ direction
of a $d$-dimensional CFT deformed by an operator respecting $z=0$ scale invariance.
Note that the dual $d$-dimensional field theory is in a flat Minkowski background, with coordinates $(u,v,x^i)$ and the reduction is
along a null direction, which would be expected to produce the standard problems and subtleties of DLCQ. In the general
case in which $f$ depends on the Sasaki-Einstein coordinates, a similar correspondence should hold.
Decomposing $f$ into harmonics of the Sasaki-Einstein, one could infer which chiral primary operators are sourced
in the dual four-dimensional conformal field theory.

From the bulk perspective, one can see immediately implications of reducing the $z=0$ geometry along a compact $u$ direction. Any asymptotically locally
anti-de Sitter geometry reduced along a spacelike circle will result in a geometry which is conformally asymptotically locally anti-de Sitter in lower
dimensions. This fact was used to analyze holography for non-conformal branes in \cite{Kanitscheider:2008kd,Kanitscheider:2009as}.
In the present case, reduction along the circle does not produce a geometry which
is conformal to anti-de Sitter, and the reason is that the reduction being carried out is not along a spacelike circle: the $u$ circle becomes null at infinity,
corresponding to the fact that $u$ is a null coordinate in the dual quantum field theory. Therefore one needs to carry out a DLCQ of the deformed theory to
obtain a Lifshitz theory in one lower dimension.

\section{Field theory analysis} \label{se:field-th}

The chiral backgrounds for general $z$ originate from deforming the dual conformal field theory by operators which respect
a non-relativistic scaling invariance. In this section we will discuss these deformations in more detail from the field theory perspective.

Consider first a conformal field theory in $d$ spacetime dimensions, with coordinates $(u,v,x^i)$. The conformal group $SO(2,d)$
contains the group of non-relativistic conformal symmetries with arbitrary $z$, which we will denote $S_{z}$.
The embedding is the following. Choosing lightcone coordinates
$u,v$, the relativistic momentum generators $P_u$ and $P_v$ are identified with ${\cal H}$ and ${\cal M}$,
respectively, the non-relativistic scaling generator ${\cal D}$ is a linear combination of the
relativistic scaling generator and a boost in the $uv$ direction and ${\cal C}$. Translations, rotations and Galilean boosts and related to translations and
rotations in the relativistic theory. More details can be found, for example, in \cite{Son:2008ye} or \cite{Maldacena:2008wh}. Note in particular
that the non-relativistic dilatation ${\cal D}$ is given in terms of the relativistic $D$ and the boost $M_{uv}$ (normalized so that the eigenvalues of $(u,v)$
are $(1, 1)$ respectively) as
\be \label{shift}
{\cal D} = D + (z-1) M_{uv}.
\ee
Thus any conformal field theory also admits the non-relativistic symmetry $S_z$.

\subsection{Marginal deformations respecting anisotropic scaling symmetry}

One can next pose the question as to what deformations preserve $S_z$ but break the relativistic conformal symmetry. Such deformations should be marginal
with respect to $S_z$, and thus the non-relativistic scaling dimension of the deforming operator should be $\Delta_{\cal D} = d$. The deforming operator
should also break Lorentz invariance, by breaking rotational symmetry in the $(uv)$ plane. The simplest possibility is a vector
operator ${\cal V}_{\mu}$ of relativistic scaling dimension $\Delta = d + (z-1)$. Using (\ref{shift}) we note that
\be
\Delta_{\cal D} ({\cal V}_v) = d; \qquad \Delta_{\cal D} ( {\cal V}_{u}) = d + 2 (z-1),
\ee
and thus ${\cal V}_v$ is marginal with respect to the non-relativistic symmetry. It is this case which is modeled holographically by gravity coupled
to massive vector fields,
\be
S_{\rm CFT} \rightarrow S_{\rm CFT} + b \int d^dx {\cal V}_{v} + \cdots
\ee
In the specific case of two dimensions,
the dual two-dimensional CFT is deformed by the right-moving component of a vector operator, namely
${\cal V}_{(1+z/2,z/2)}$ with holomorphic and anti-holomorphic dimensions $(h_v,h_u) = (1+z/2,z/2)$, so that
\be
S_{\rm CFT} \rightarrow S_{\rm CFT} + b \int dv du {\cal V}_{(1+ z/2,z/2)} + \cdots
\ee
with $b$ constant and where the ellipses denote terms higher order in $b$. This deformation is manifestly consistent with the non-relativistic scaling symmetry
\be
v \rightarrow \lambda^{2-z} v; \qquad
u \rightarrow \lambda^{z} u,
\ee
along with translational symmetries in the $(u,v)$ direction. Note that the combination:
\be
\chi^2 \equiv v^{z} u^{z-2} \label{x-def}
\ee
is invariant under the non-relativistic scaling symmetry, whilst the (Lorentz-invariant) combination $(2 u v)$ scales as $\lambda^2$.

It is interesting to note that such deformations by vector operators are only one special case of a more general situation in two dimensions, in which one
deforms a 2d CFT by a $(p,q)$ operator ${\cal Y}_{p,q}$ where $(p,q)$ are the CFT scaling weights corresponding to $(v,u)$ respectively,
\be
S_{\rm CFT} \rightarrow S_{\rm CFT} + b_{p,q} \int d^2x {\cal Y}_{p,q}.
\ee
As discussed in \cite{Guica:2010sw} such a deformation respects anisotropic scale invariance with exponent $z$ under which $u \rightarrow \lambda^{z} u$
and $v \rightarrow \lambda^{2-z} v$ provided that
\be
(p-1) (z-2) = (q-1) z. \label{scal-form}
\ee
Vector deformations in which $p = q \pm 1$ are just one special case. Another interesting case is that of strictly
chiral deformations of conformal field theories, by which we mean
\be
S_{\rm CFT} \rightarrow S_{\rm CFT} + b_{p,0} \int d^2 x {\cal Y}_{p,0},
\ee
where ${\cal Y}_{p,0}$ is a holomorphic field of arbitrary integral spin. The dynamical exponent in this example is
\be
z = 2 \left ( 1 - \frac{1}{p} \right ).
\ee
The case of $p=1$ corresponds to deformation by a conserved current, which as we saw earlier is trivial from the bulk
perspective; that of $p=2$ corresponds to $z=1$ anisotropic symmetry and could be realized by deforming with the holomorphic component
of the stress energy tensor. Such chiral deformations of CFTs have arisen previously in many contexts, from two-dimensional large $N$ QCD to Kodaira-Spencer theory,
see for example \cite{Dijkgraaf:1996iy}, but the existence and implications of
the anisotropic scaling symmetry have not been discussed. From the form of (\ref{scal-form}) one can see that a theory with exponent $z$ can also be viewed
as a theory with exponent $z' = (2-z)$ upon exchanging the r\^{o}les of $u$ and $v$.

Returning to the case of vector deformations,
while such deformations are manifestly marginal, one also needs to show that they are exactly marginal. A priori, one might not have
expected such deformations to be exactly marginal with respect to the non-relativistic symmetry group. However, holographic duals for
such deformations (at strong coupling) exist generically, and this implies that such operator deformations do indeed remain exactly marginal. Using conformal
perturbation theory, the correction to the two point function of the deforming operator itself,
 in the deformed theory, is expressed in terms of higher point functions in the conformal theory as
\be \label{int1}
\delta \< {\cal V}_{v} (x) {\cal V}_{v} (0) \> = \sum_{n \ge 1} \frac{1}{n!} \< {\cal V}_{v} (x) \prod_{a=1}^{n} \int dx_a (b {\cal V}_{v}(x_a)) {\cal V}_v (0) \>.
\ee
This expression can be rewritten in momentum space as
\be
\delta \< {\cal V}_{v} (k) {\cal V}_{v} (-k) \> = \sum_{n \ge 1} \frac{1}{n!} \< {\cal V}_{v} (k) (b {\cal V}_{v}(0))^n {\cal V}_v (-k) \>.
\ee
If the deformation is to be exactly marginal, at zero momentum, the anomalous dimension of the operator must vanish at zero momentum.
A simple argument why this is true was given in \cite{Guica:2010sw} for the case of $z=2$ and follows from (relativistic) conformal invariance, which implies that
\be
\< {\cal V}_{v} (k) (b {\cal V}_{v}(0))^n {\cal V}_v (-k) \> = (b k_v)^n \< {\cal V}_{v} (k) {\cal V}_{v} (-k) \> f^{(n)} \left (\ln (k^2/\mu^2) \right ),
\ee
where the function $f^{(n)}$ can depend at most logarithmically on the scale. The right-hand side always
vanishes for $k_v \rightarrow 0$, and therefore the deforming
operator itself cannot acquire an anomalous dimension. 

For general values of $z$ (excluding the cases where $z/2$ is an integer) the argument is even simpler because
the integrals appearing in \eqref{int1} are not scale invariant. This implies, following section 4.4 of 
\cite{Guica:2010sw}, that for generic values of $z$ no operators acquire anomalous scaling dimensions in the deformed theory (again, except
when $z/2$ is an integer). Instead the corrections to the two point function of the deforming operator are simply of the form
\be
\< {\cal V}_{v} (k) (b {\cal V}_{v}(0))^n {\cal V}_v (-k) \> = (b k^{z/2}_v k_u^{z/2 - 1})^n
\< {\cal V}_{v} (k) {\cal V}_{v} (-k) \>, 
\ee
where no logarithmic term can appear on the right hand side. The quantity $(k_{v}^{z} k_{u}^{z/2-1})$ is, according to \eqref{x-def},
invariant under the anisotropic scaling symmetry and therefore the deformation corrects only the normalization of the operator but not its non-relativistic 
scaling dimension. Thus
the operator indeed remains marginal in the deformed anisotropic theory.  

Note that an analogous simple argument cannot be made for deformations by marginal scalar operators. In such a case the deformation of the scalar two point
function is
\be
\delta \< {\cal O} (x) {\cal O} (0) \> = \sum_{n \ge 1} \frac{1}{n!} \< {\cal O} (x) \prod_{a=1}^{n} \int dx_a (\alpha {\cal O}(x_a)) {\cal O} (0) \>,
\ee
where $\alpha$ is a scalar parameter. Conformal invariance implies that
\be
\< {\cal O} (k) (\alpha {\cal O}(0))^n {\cal O} (-k) \> = \alpha^n \< {\cal O} (k) {\cal O} (-k) \> f^{(n)} \left (\ln (k^2/\mu^2) \right ),
\ee
and if any of the $f^{(n)}$ are non-zero the operator acquires an anomalous dimension. Generically the $f^{(n)}$ are indeed non-zero, and
one needs to use additional structure such as supersymmetry to determine when operators are exactly marginal.

\subsection{Deformations with $z=0$}

Scaling symmetry with $z=0$ cannot be realized non-trivially with by vector operator deformations. The vector operator which would
respect $z=0$ has relativistic dimension $(d-1)$ and is a conserved current. The deformation by a constant null source for this operator
introduces chemical potentials in the dual theory and breaks the relativistic invariance in a trivial way; correspondingly the bulk metric
remains $AdS_{d+1}$ after the deformation. In section \ref{z0} we showed that $z=0$ bulk solutions could be obtained by coupling gravity
to a massless scalar, and switching on a profile for the scalar field which depends on the lightcone coordinate $u$. Let us now discuss
the corresponding field theory deformations.

Working to leading order around the $AdS_{d+1}$ background, the solution \eqref{non-rel-z0} corresponds to a deformation of the CFT,
\be
S_{\rm CFT} \rightarrow S_{\rm CFT} + \int du F(u) \int dv d^{d-2} x {\cal O}_{d},
\ee
where the operator ${\cal O}_d$ is a marginal scalar operator dual to the bulk field $\Phi$. Recalling that the scaling symmetry with $z=0$ acts
as
\be
u \rightarrow \lambda^{0} u; \qquad
v \rightarrow \lambda^2 v; \qquad
x \rightarrow \lambda x,
\ee
and that the marginal scalar operator with scale as ${\cal O}_{d} \rightarrow \lambda^{-d} {\cal O}_d$, one notes that the deformation indeed
respects $z=0$ symmetry for any choice of the function $F(u)$. The question next arises as to whether this deformation is exactly marginal, since
as we discussed above, marginal scalar operators are generically not exactly marginal. However, in the bulk realization, the scalar operator is
a chiral primary which is exactly marginal. In the holographic realizations, therefore, the deformation
is indeed exactly marginal for any choice of $F(u)$, with the case of constant $F(u)$ being a special case in which relativistic symmetry remains unbroken.

\subsection{Correlation functions in the deformed theory}

Next let us consider the behavior of correlation functions under such deformations, focusing on the case of two dimensions.
Suppose that in the original CFT the stress energy tensor is $T_{ab}$, the vector operator of relativistic dimension $(1+z)$ is $V_a$ and let
${\cal O}_{h,\bar{h}}$ be generic chiral operators of relativistic dimension $(h,\bar{h})$. Here $v$ corresponds to the holomorphic
coordinate, scaling weight $h$, and $u$ corresponds to the anti-holomorphic coordinate, scaling weight $\bar{h}$.
The corresponding non-relativistic scaling dimension for the operator ${\cal O}_{h, \bar{h}}$ is
\be
{\Delta}_{{\cal D}} = h (2 -z) + \bar{h} z. \label{por}
\ee
Non-relativistic scale invariance generically constrains the two point functions to be of the form
\be
\< {\cal O}_{\Delta_{\cal D}} (u,v) {\cal O}_{\Delta_{\cal D}'} (0) \> =
\frac{1}{u^{(\Delta_{\cal D} + \Delta_{\cal D}')/z}} f (\chi), \label{gfr}
\ee
where $f(\chi)$ is an arbitrary function of the scale invariant quantity $\chi$ defined in (\ref{x-def}).
The relativistic two-point function is of the required form noting that
\be
\frac{1}{v^{2h} {u}^{2 \bar{h}}} \equiv \frac{1}{{u}^{2 {\Delta}_{\rm nr}/z}} {\chi}^{-2h/z} \equiv \frac{1}{ (uv)^{\Delta_{\rm nr}}}
{\chi}^{\bar{h} - h}.
\ee
One should note that for generic $z$ operators of different scaling dimension can have non-vanishing two point functions.
In the cases of $z=1$ and $z=2$ the additional special conformal symmetry imposes the further restriction that $\Delta_{\rm nr} = \Delta_{\rm nr}'$.

Using conformal perturbation theory one can derive the corrections to the two point function at non-zero $b$. To leading order this results in
(see section 4.4 of \cite{Guica:2010sw} for a detailed analysis),
\be
\< {\cal O} (u,v) {\cal O} (0) \> = \frac{1}{v^{2h} {u}^{2 \bar{h}}} \left(
c_{0} + c_{1} b (\chi)^{-1/2} \right ), \label{1st}
\ee
where $c_{0}$ denotes the operator normalization in the CFT and $c_1$ is a computable numerical constant, proportional to the structure constant
of the three point function between these operators and the deforming vector operator. When $(z/2)$ is an integer the
corresponding expression involves logarithms and is instead of the form,
\be
\< {\cal O} (u, v) {\cal O} (0) \> = \frac{1}{v^{2h} {u}^{2 \bar{h}}} \left(
c_{0} + c_{1} b \chi^{-1/2} \ln( m^2 (u v)) \right ).
\ee
The appearance of logarithms reflects the fact that operators acquire anomalous scaling dimensions in the deformed theory; this is only
possible when $z/2$ is an integer. 

Returning to the generic case where $z/2$ is not an integer, 
the corrections are organized in powers of $b {\chi}^{-1/2} $ since the deformed action remains invariant
under the original dilatation symmetries provided that the coupling $b$ is also transformed. Working to higher orders in $b$ in the case
where $z/2$ is non-integral gives
\be
\< {\cal O} (u, v) {\cal O} (0) \> = \frac{1}{v^{2h} {u}^{2 \bar{h}}} \sum_{n} c_n b^n (\chi)^{-n/2}
\ee
The corrections in $b$ hence change the ${\chi}$ dependent normalization of the correlator, but do not the scaling dimension of the operator. By contrast in
the case where $z/2$ is integral the logarithmic terms in the expansion in $b$ indicate that the scaling dimension is also modified at non-zero $b$; the case of
Schr\"{o}dinger symmetry, $z=2$, was the main focus of \cite{Guica:2010sw}.

In the holographic realizations considered here, there are no three-point couplings between the metric and the vector field in the bulk action. This implies that the
leading corrections to their two point functions occur at order $b^2$, and they are related to four point functions at the conformal point. More generally, since
all odd couplings vanish in the bulk, their corrected two point functions involve functions of $(b^2/ \chi)$.  For generic $z$
the stress energy tensor and the vector operator can have non-zero two point functions with each other, at non-zero $b$, and indeed as we will show the
Ward identities do force these two point functions to be non-zero.

\subsection{Counterterms and renormalizability} \label{counter}

In this section we will consider what counterterms are needed in computing the two point functions in conformal perturbation theory. Explicit expressions
for the corrections to correlation functions at leading order in $b$ were obtained in \cite{Guica:2010sw} using the method of differential regularization
\cite{Freedman:1991tk}. Counterterms in this method are implicit, although they can be constructed explicitly as in \cite{Freedman:1992gr}. In the case at hand
we would like to explore the structure of the required counterterms and compare it with the counterterms obtained in holographic renormalization.

Following \cite{Guica:2010sw}, the leading order correction \eqref{1st} is determined by the three point function between the deforming operator
and the other two operators. Analytically continuing to Lorentzian signature via $v \rightarrow w$ and $u \rightarrow \bar{w}$ the correction behaves as
\bea
\delta \< {\cal O} (w, \bar{w}) {\cal O} (0) \> & \sim &
\frac{b}{w^{2h - 1 - z/2} {\bar{w}}^{2 \bar{h} - z/2}} \int
\frac{d^2y}{(w-y)^{z/2+1} y^{z/2+1} (\bar{w} - \bar{y})^{z/2} \bar{y}^{z/2}}; \label{3pf2} \\
& \sim & \frac{b}{w^{2h - 1 - z/2} {\bar{w}}^{2 \bar{h} - z/2}} \pa_w^2 \int
\frac{d^2 y}{|y - w|^z |y|^z}. \label{3pf}
\eea
When $2z$ is not an integer, then $|y|^{-z}$
is well-defined as a distribution, and its Fourier transform is
\be
\int dw d \bar{w} e^{- i  k w - i \bar{k } \bar{w}} |w|^{-z} = \pi 2^{2-z} \frac{\Gamma (1-z/2)}{\Gamma(z/2)} |k|^{z - 2}.
\ee
Noticing that the integral \eqref{3pf} is a convolution, the integral may be computed via the inverse Fourier transform of the products
of the two Fourier transforms. This results in a leading correction to the two point function of the form \eqref{1st}.

Whilst the correct finite contribution to the two point function is obtained in this way, note that the integrals being computed are in general divergent
and additional counterterms are required relative to $b=0$. One way
to see this is remove small circles of radius $\Lambda^{-1}$ around points where the vertices coincide; with this regulator the integral will have power
divergences which can be cancelled by adding contact terms. Let us consider the case where the operator is
the deforming operator itself. The new counterterms are then precisely the same counterterms
needed in computing the three point function of this (for $z > 1$, irrelevant) operator in the CFT.
The counterterms at order $b^n$ will similarly be related to the counterterms
that arise in computing $(n+2)$-point functions, and the latter must on general grounds be local, covariant functionals of the vector operator sources.

Let us restrict to the case where the conformal field theory is treated within the flat background.
The leading order counterterms at $2n$-th order in the vector field sources diverge as
\be
S_{\rm ct} \sim \Lambda^{2 n (z - 1)  + 2} \int d^2 x (b^a b_a)^n + \cdots, \label{ct1}
\ee
and we have used the fact that the counterterm is necessarily covariant. The degree of divergence is computed
using the known dimensionality of the operator source, of the metric and of derivatives.
Since counterterms must be scalars, any additional
derivatives acting on the sources will reduce the degree of divergence,
\be
S_{\rm ct} \sim \Lambda^{2 n (z - 1) + 2 - 2m} \int {t}^{a_1 \cdots a_{2n} c_1 \cdots c_{2m}} \prod^{2m}_{i} D_{c_i}
\prod^{2n}_{j} b_{a_j}, \label{ct2}
\ee
where ${t}^{a_1 \cdots a_{2n} c_1 \cdots c_{2m}}$ is a tensor, which must include $(m+n)$ inverse metrics, since
the counterterm is a scalar. Compared to \eqref{ct1}, these terms are indeed more divergent when $ m > 0$. The actual tensors which
arise need to be obtained by explicit computation, but note that when $m=0$ the tensor needs to be completely symmetric and built
out of the (flat) metric, with \eqref{ct1} being the only possibility.

Now let us suppose one has computed the counterterms to arbitrarily high order in the vector field sources and then let
\be
b_a = b \delta_{au} + a_{a},
\ee
where $b$ is constant and finite whilst $a_a$ is treated perturbatively. The $2n$-point correlation functions in the deformed
theory may then be computed by working to order $2n$ in the source $a_a$. Consider which counterterms can contribute to this calculation: when $a_a =0$,
all of the counterterms vanish, since there are no covariant scalar invariants which can be formed from a null vector field. The absence of such scalar
invariants is related to the exactly marginal nature of the anisotropic deformation.

In computing the two point functions in the deformed theory, one needs to retain only terms quadratic in the source $a_a$. For $z < 2$ this implies
that only a finite number of counterterms are needed. This follows from \eqref{ct2}: since the background source is null, at order $n$ we need to
include at least $m = (n-2)$ $v$ derivatives to form a scalar invariant. The leading non-vanishing counterterms have the structure
\be
S_{\rm ct} \sim \Lambda^{2 n (z-2) + 6 - 2 \td{m}} \int d^2 k k^{2 \td{m}} (b k_v)^{2n-4} (b a_v)^2,
\ee
where we work in momentum space, $k_v$ is the lightcone momentum and $k$ schematically denotes all momenta. Clearly for $z < 2$ there are only a finite
number of divergent counterterms. However, for $z > 2$, counterterms of arbitrarily high order in the finite source $b$ can contribute.
In the holographic computation we will find the same analytic structure, and we will argue in addition that for non-rational values of $z$ the counterterms
cannot give finite contributions to the renormalized two point functions.

\subsection{Stress energy tensor and deforming vector operator} \label{field-ward}

Let us next consider the stress energy tensor and the deforming vector operator, focusing on the case of $z <1$ where the latter is a
relevant operator. Our starting point is a two-dimensional conformal field theory which is invariant under diffeomorphisms and Weyl
rescalings (up to the usual conformal anomaly). If the generating functional of
the field theory is $W$ the stress energy tensor ${\cal T}_{ab}$ may be defined as\footnote{Note that we are working here in Lorentzian signature.}:
\be
{\cal T}_{ab} = \frac{2i}{\sqrt{-g_{(0)}}} \frac{\delta W}{\delta g_{(0)}^{ab}},
\ee
where $g_{(0) ab}$ is the background metric for the field theory. The vector operator ${\cal V}_a$ of scaling
dimension $(1 + z)$ which couples to a source $b^{a}$ is correspondingly defined as:
\be
{\cal V}_a = \frac{i}{\sqrt{-g_{(0)}}} \frac{\delta W}{\delta b^a}.
\ee
Diffeomorphisms act as
\be
\delta g_{(0)}^{ab} = - (D^a \zeta^b + D^b \zeta^a); \qquad
\delta b_{a} = \zeta^c D_{c} b_a + D_a \zeta^c b_c, \label{diffeo}
\ee
with $D_a$ the covariant derivative. Weyl transformations act as
\be
\delta g_{(0)}^{ab} = - 2 \lambda g_{(0)}^{ab}, \qquad \delta b_a = z \lambda b_a.
\ee
Imposing invariance of the generating functional under diffeomorphisms and Weyl transformations gives the following Ward identities:
\bea
D^{b} \< {\cal T}_{ab} \>_J - b_a D^{b} \< {\cal V}_b \>_J + F_{b a} \< {\cal V}^b \>_J = 0; \\
\< {\cal T}^{a}_a \>_J - z b^a \< {\cal V}_a \>_{J} = {\cal A} [g_{(0)},b].
\eea
where $\< {\cal O} \>_J$ denotes the expectation value of an operator ${\cal O}$ in the presence of sources $J$, and ${\cal A}$
denotes the conformal anomaly. Here $F_{ab}$ is the curvature of the vector field source $b_a$, $F_{ab} = 2 \pa_{[a} b_{b]}$.
The anomaly can in principle depend covariantly both on the background metric $g_{(0)}$ and on the source $b$. Since the anomaly must
have dimension two, for generic values of $z$ there is no covariant quantity of the right weight that can be formed out of $b$
and the anomaly will be
given entirely in terms of the Ricci scalar of the metric $g_{(0)}$ and the central charge $c$ of the CFT:
\be
{\cal A} (g_{(0)}) = \frac{c}{24 \pi} R[g_{(0)}]
\ee
(The additional factor of  $2 \pi$ on the righthand side relative to usual CFT conventions follows
from the absence of the $2 \pi$ factor in our normalization of the stress energy tensor.) For specific values of $z$ where a quantity of
the form $\pa^n b^p$ can dimension two there are additional contributions to the conformal anomaly, as will be discussed in section \ref{sec:hr}.

The Ward identities imply an infinite number of relations for correlation functions in the deformed theory, which are obtained by
differentiating with respect to the sources and then setting $g_{(0) ab}$ and $b_a$ to their background values. In particular, the
identities for two point functions can be completely solved, up to the two point functions of the vector operators. For notational convenience
let us denote
$T \equiv T_{vv}$, $\bar{T} \equiv T_{uu}$ and $\theta \equiv T_{uv}$. The dilatation Ward identity implies that
\be
\< \theta (u,v) {\cal V}_v (0) \> = \half z b \< {\cal V}_v (u, v) {\cal V}_v (0) \>; \qquad
\< \theta (u,v) {\cal V}_u (0) \> = \half z b \< {\cal V}_v (u, v) {\cal V}_{u} (0) \>,
\ee
whilst
\be
\< \theta (u,v) \theta (0) \> = \frac{1}{4} z^2 b^2 \< {\cal V}_v (u,v) {\cal V}_v (0) \> + \cdots,
\ee
where the ellipses denote local terms. Solving the $v$ component of the diffeomorphism identity then results in
\bea
\< T(u,v) T(0) \> &=& \frac{c}{8 \pi^2 v^4} + \frac{1}{4} z^2 b^2 \frac{\pa_v^2}{\pa_{u}^2} \left (\< {\cal V}_v (u,v) {\cal V}_v (0) \> \right ); \nn \\
\< T(u,v) \theta (0) \> &=& - \frac{1}{4} z^2 b^2 \frac{\pa_v}{\pa_{u}} \left ( \< {\cal V}_v (u,v) {\cal V}_v (0) \> \right ). \\
\< T(u,v) \bar{T}(0) \> &=& \frac{b^2 z}{4} (z-2)  \< {\cal V}_v (u,v) {\cal V}_v (0) \>  - \frac{b^2 z^2}{2} \frac{\pa_{v}}{\pa_u}
 \left ( \< {\cal V}_u (u,v) {\cal V}_v (0) \> \right ); \nn \\
\< T(u,v) {\cal V}_v (0) \> &=& - \half z b  \frac{\pa_v}{\pa_{u}} \left ( \< {\cal V}_v (u,v) {\cal V}_v (0) \> \right ); \nn \\
\< T(u,v) {\cal V}_u (0) \> &=& - \half z b  \frac{\pa_v}{\pa_{u}} \left ( \< {\cal V}_u (u,v) {\cal V}_v (0) \> \right ), \nn
\eea
where local terms have been suppressed. Real-time issues and contact terms in the correlators have also been suppressed, since they
do not play a r\^{o}le in the discussions here.
Solving the $u$ component of the diffeomorphism identity results in
\bea
\< \bar{T}(u,v) \bar{T}(0) \> &=& \frac{c}{8 \pi^2 {u}^4} + b^2 (2-z) \left ( \frac{\pa_u^2}{2 \pa_v^2}
\< {\cal V}_v (u,v) {\cal V}_v (0) \> + \frac{\pa_u}{\pa_v} \< {\cal V}_v (u,v) {\cal V}_u (0) \> \right ) \nn \\
&& \qquad + b^2 \< {\cal V}_u (u,v) {\cal V}_u (0) \>; \nn \\
\< \bar{T}(u,v) \theta (0) \> &=& \frac{1}{4} z b^2 (2 - z) \frac{\pa_u}{\pa_{v}} \left ( \< {\cal V}_v (u,v) {\cal V}_v (0) \> \right )
+ \frac{b^2 z}{2}  \< {\cal V}_u (u,v) {\cal V}_v (0) \>; \\
\< \bar{T}(u,v) {\cal V}_v (0) \> &=& b (1 - \half z)   \frac{\pa_u}{\pa_{v}} \left ( \< {\cal V}_v (u,v) {\cal V}_v (0) \> \right )
+ b  \< {\cal V}_u (u,v) {\cal V}_v (0) \> ; \nn \\
\< \bar{T}(u,v) {\cal V}_u (0) \> &=& b (1 - \half z)  \frac{\pa_u}{\pa_{v}}
\left ( \< {\cal V}_u (u,v) {\cal V}_v (0) \> \right ) + b \< {\cal V}_u (u,v) {\cal V}_u (0) \>. \nn
\eea
Thus in the deformed theory the relativistic stress energy tensor is no longer conserved and has non-trivial two point functions, determined
in terms of the conformal anomaly of the original theory and the correlation functions of the vector operator.

\bigskip

In the deformed theory the relativistic stress energy tensor is no longer conserved. However,
when $b$ is covariantly constant and has zero curvature $F \equiv 0$, there is a non-symmetric stress-energy tensor $t_{ab}$ defined such that
\be
t_{ab} = {\cal T}_{ab} - b_a {\cal V}_b, \label{non}
\ee
which is covariantly conserved. As discussed in \cite{Guica:2010sw}, the components of $t_{ab}$ are related to the Noether
charges, including those associated with the translational symmetries $u \rightarrow u  +a$, $v \rightarrow v + b$, and the fact
that $t_{ab}$ is non-symmetric follows from the general result that a conserved stress tensor in any field
theory in a Minkowski background which breaks Lorentz invariance cannot be symmetric.

The conserved stress energy tensor $t_{ab}$ is obtained by coupling the CFT to a vielbein $e^{a}_{\hat{a}}$, and then defining
\be
t_{a b} = e_{b}^{\hat{b}} \frac{i}{|e|} \frac{\delta W_{\rm cft}}{\delta e^{a \hat{b}}},
\ee
where $\hat{a}$ denotes tangent space indices. The vector operator deformation is then given by
\be
S_{\rm cft} \rightarrow S_{\rm cft} + i \int d^2 x |e| b^a e^{\hat{a}}_{a} {\cal V}_{\hat{a}}.
\ee
If we now consider the behavior of the generating functional under Lorentz transformations, diffeomorphisms and Weyl transformations,
respectively, we can derive the following Ward identities for $t_{ab}$:
\bea
\< t_{[ab]} \> + b_{[a} \< {\cal V}_{b]} \> &=& 0; \\
D^{\hat{a}} \< t_{\hat{b} \hat{a}} \> + D^{b} b_{\hat{b}} \< {\cal V}_b \> + F_{b \hat{b}} \< {\cal V}^b \> &=& 0; \nn \\
\< {t}^{a}_a \> + (1 - z) b^a \< {\cal V}_a \>_{J} &=& {\cal A} (e), \nn
\eea
where we again assume that the only anomaly is the conformal anomaly ${\cal A} (e)$, which depends only on the scalar curvature. The operator
$t_{ab}$ is indeed conserved when $b$ is covariantly constant and has zero curvature.

For $z < 1$ deformations, which are relevant with respect to the conformal symmetry group, both the relativistic stress energy tensor ${\cal T}_{ab}$
and the (conserved) anisotropic stress energy tensor $t_{ab}$ are natural well-defined operators to consider. The relativistic stress energy tensor
${\cal T}_{ab}$ is natural when we are treating the theory as a deformation of a CFT, whilst the tensor $t_{ab}$ is natural if we view the theory as
an intrinsically anisotropic scale-invariant theory, which acquires additional symmetries in the UV.
Correlation functions of the tensors are non-locally
related to each other, but may be obtained straightforwardly by the defining relation (\ref{non}).
For $z > 1$ deformations, however, which are irrelevant with respect to the conformal symmetry, it is rather less natural to work with the operator ${\cal T}_{ab}$,
since it is neither conserved nor is the theory conformal in the UV. However, for generic values of $z$ such that
$z > 1$, two point functions around the scale invariant vacuum,
including correlation functions of ${\cal T}_{ab}$, are reconstructable from those of the deforming vector operator, using the Ward identities, and we thus evade
having to work in a vielbein formalism.

\section{Holographic renormalization for $d=2$} \label{sec:hr}

In this section we will derive general expressions for the renormalized holographic one point functions of dual operators
in terms of coefficients in the near boundary expansions of bulk solutions. We will focus first on the case of $z < 1$ in two dimensions,
and then comment on the case of $z > 1$ which is no longer asymptotically AdS. We discuss holographic
renormalization using two methods. The first uses the general asymptotic solution of the bulk field equations to regulate the volume
divergences of the on-shell action with covariant counterterms being obtained by inverting these expansions. This method is the most familiar
approach to holographic renormalization, see the review \cite{Skenderis:2002wp},
but becomes increasingly cumbersome as the number of counterterms required increases. Since this method
works with the asymptotic expansion of the bulk metric and vector, it allows us to appreciate the roles of different terms in the asymptotic expansions,
which we will exploit in section \ref{se:linear}.

The second method of holographic renormalization
exploits the Hamiltonian approach developed in \cite{Papadimitriou:2004ap,Papadimitriou:2005ii},
which uses covariant expansions in terms of eigenfunctions of a dilatation operator. This approach is much more efficient
and powerful; the main advantage here is that renormalized correlation functions can, in favorable cases, be determined without explicit computation of
the counterterms. In cases where many counterterms are needed, and the inversion of the asymptotic expansions of the bulk fields is cumbersome, this methodology
is the more appropriate one to use.

\subsection{Asymptotic expansions and their inversion}

We begin by analyzing the most generally asymptotically locally AdS solutions of the bulk field equations.
In the neighborhood of the conformal boundary at $\rho \rightarrow 0$, the metric and vector field can be expressed as:
\bea
ds^2 &=& \frac{d \rho^2}{4 \rho^2} + \frac{1}{\rho} g_{ab} (x,\rho) dx^{a} dx^{b}; \label{fg-form} \\
B_{a} &=& \rho^{-z/2} b_{a} (x, \rho); \nn \\
B_{\rho} &=& \rho^{m} b_{\rho} (x,\rho), \nn
\eea
where the power $m$ will be determined below.
Given this coordinate choice, the Einstein equations can be written as:
\bea
&& R_{ab} [g_{cd}] + (d-2) g'_{ab} + {\rm{tr}} (g^{-1} g') g_{ab} - \rho (2 g''
- 2 g' g^{-1} g' + {\rm{tr}}
(g^{-1} g') g')_{ab} = \frac{1}{2} m^2 B_{a} B_{b} \label{f-eq} \\
&& \qquad - \frac{1}{4 (d-1)} \rho ( g^{ce} g^{df} F_{cd} F_{ef} + 8 \rho g^{cd}
 F_{c \rho} F_{d \rho}
) g_{ab} + \frac{1}{2} \rho ( g^{cd} F_{ac} F_{bd} + 4 \rho F_{a \rho} F_{b \rho
}); \nn \\
&& D_{a} ( {\rm tr} (g^{-1} g')) - D^{b} g'_{ab} = - (m^2 B_a B_{\rho} + \rho g^
{cd} F_{ac}
F_{\rho d} ); \nn \\
&& \frac{1}{4} {\rm tr} (g^{-1} g' g^{-1} g') - \frac{1}{2} {\rm tr} (g^{-1} g''
) =
\frac{1}{2} m^2 B_{\rho} B_{\rho} - \frac{1}{16 (d-1)} g^{ce} g^{df} F_{cd} F_{e
f} +
\frac{(d-2)}{2 (d-1)} \rho g^{cd} F_{c \rho} F_{d \rho}. \nn
\eea
The vector field equations in this coordinate system become:
\bea
\pa_{a} \left ( \sqrt{-g} g^{ab} F_{b \rho} \right ) &=& \frac{m^2}{\rho} \sqrt{
-g} B_{\rho}; \\
\pa_a \left ( \sqrt{-g} g^{ab} g^{cd} F_{bd} \right ) + 4 \rho^{d/2 - 1}
\pa_{\rho} \left ( \sqrt{-g} \rho^{2 - d/2} g^{cd} F_{\rho d} \right ) &=& \frac
{m^2}{\rho} \sqrt{-g}
g^{cd} B_{d}. \nn
\eea
The divergence equation for the vector field is:
\be \label{div}
\pa_{a} \left ( \sqrt{-g} g^{ab} B_{b} \right ) + 4 \rho^{d/2} \pa_{\rho}
\left ( \sqrt{-g} \rho^{1-d/2} B_{\rho} \right ) = 0.
\ee
Here for future convenience the equations are written for general $d$, although in this section we will consider only $d=2$.
The leading order terms in these equations as $\rho \rightarrow 0$  imply that:
\be
g_{ab}(x,0) = g_{(0) ab} (x); \qquad
b_{a} (x,0) = b_{(- z)a} (x),
\ee
for arbitrary (non-degenerate) metric and 1-form respectively. By the usual rules of AdS/CFT, $g_{(0)}$ acts as a source for
the stress energy tensor in the dual theory, whilst $b_{(-z) a}$ acts a source for the dual vector operator of dimension $(d + z -1)$.

The leading term in the expansion of $B_{\rho}$ (including the polynomial power $m$)
is determined by the divergence equation, and does not therefore represent an additional independent source. Indeed,
using the leading order terms in the equations of motion one finds
that the power of $\rho$ in the leading order term of $B_{\rho}$ is the same
as in the leading term of $B_{a}$:
\bea
B_{\rho} &=& \rho^{-z/2} \left [ b_{(-z)\rho} + \cdots \right ]; \\
b_{(-z)\rho} &=&  \frac{1}{2z} D_{(0)}^{a} b_{(-z) a}, \nn
\eea
where $D_{(0)}$ is the covariant derivative associated with the metric $g_{(0) ab}$.
Therefore the value of $m$ in (\ref{fg-form}) is $-z/2$.

The first step in holographic renormalization
is to determine the general asymptotic expansion near the boundary, namely the radial expansion of the fields. We thus expand the fields in
Fefferman-Graham form as:
\bea
g_{ab}(x, \rho) &=& g_{(0) ab} (x) + \cdots + \rho g_{(2) ab} (x) + \cdots; \\
b_{a} (x, \rho) &=& b_{(-z) a} (x) + \cdots + \rho^{z} b_{(z) a} (x) +  \cdots; \nn \\
b_{\rho} (x,\rho) &=&   b_{(-z) \rho} (x) +\cdots + \rho^{z} b_{(z) \rho} (x) + \cdots. \nn
\eea
The radial expansion only needs to
be calculated to sufficient order to determine the divergences in the on-shell action; in practice this means up to the order at which coefficients
are undetermined or only partially determined by the asymptotic analysis.Since the coefficients in the field equations
(\ref{f-eq}) are polynomials in $\rho$
this system of equations admits solutions with $(g_{ab} (x,\rho),  b_{a} (x, \rho), b_{\rho} (x,\rho))$  regular functions of $\rho$.
To solve these equations, one may successively differentiate the equations w.r.t. $\rho$ and set $\rho = 0$. In pure gravity, the metric
is expanded in integral powers of $\rho$, with additional logarithmic terms generically needed to solve the equations of motion in odd dimensions.
In the case of gravity coupled to the massive vector, the powers of
$\rho$ that occur in the expansions need to be determined from the equations of motion, and should not a priori be assumed to be integral.

Explicit solution of the equations of motion for $0 < z < 1/2$ determines that the first subleading term in the metric is
actually of order $\rho^{1-z}$.
The form of the asymptotic expansions for $0 < z < 1/2$ can be summarized as follows. The only terms required in
determining the counterterms and renormalized one point functions are
\bea
g_{ab} &=& g_{(0)ab} + \rho^{1-z} g_{(2- 2z)ab} + \rho g_{(2)ab} + \tilde{h}_{(2)ab} \rho \log\rho + \cdots; \label{exp1} \\
b_{a} &=& b_{(-z)a} + \rho^{z} b_{(z)a} + \cdots; \nn \\
b_{\rho} &=& b_{(-z)\rho} + \rho^{z} b_{(z)\rho} + \cdots, \nn
\eea
where the ellipses denote subleading terms.
The following coefficients are completely determined in terms of the non-normalizable modes:
\bea
g_{(2-2 z) ab} &=& \frac{z}{2(1-z)} \left [ b_{(-z) a} b_{(-z)  b} - \half {\rm {Tr}} (b_{(-z)} g_{(0)}^{-1} b_{(-z)}) g_{(0) ab} \right ], \\
b_{(-z)\rho} &=& \frac{1}{2z} D_{(0)}^{a} b_{(-z)a}, \nn \\
\tilde{h}_{(2)ab} &=& \half \left ( R_{(0)ab} - \half g_{(0)ab} R_{(0)} \right) = 0. \nn
\eea
In the latter expression the identity relating Ricci curvature and Ricci scalar in two dimensions has been imposed.

In the vector field, the coefficient $b_{(z) a}$ is totally undetermined, whilst
\be
b_{(z)\rho} = - \frac{1}{2z} D_{(0) a} b_{(z)}^{a}.
\ee
The metric coefficient $g_{(2) ab}$ is undetermined, but subject to the following constraints:
\bea
{\rm {Tr}} g_{(2)}
&=& - \half R_{(0)} + z^{2} {\rm Tr} (b_{(-z)}g_{(0)}^{-1} b_{(z)}); \\
D_{(0)}^{b} g_{(2) ba}  &=&
\partial_{a} \left ( {\rm {Tr}}(g_{(2)}) \right ) + \frac{z}{2} \left ( b_{(-z) a} b_{ (z) \rho} + b_{(-z) \rho} b_{(z) a} \right ) \nn \\
&& \qquad + \frac{z}{2} \left (
F_{(-z) ac} b_{(z)}^{c} - F_{(z) ac} b_{(-z)}^c \right ), \nn
\eea
where $F_{(-z) ab}$ is the curvature of the field $b_{(-z) a}$.

For generic $0 < z < 1$, the asymptotic expansion of the metric has the form
\be
g_{ab} = \sum_{m,n} g_{(2m + 2n (1-z))} \rho^{m + n(1-z)} + \cdots \label{exp2}
\ee
with $(m,n)$ integral and coefficients of terms with
\be
m + n(1-z) < 1 \label{exp3}
\ee
contribute to the on-shell divergences. For $1/2 < z < 1$ this implies that an increasing number of coefficients can contribute
to the on-shell divergences, and the Hamiltonian approach to renormalization is more efficient. Note also that
the coefficient $g_{2n(1-z)}$ is of order $b_{-z}^{2n}$, and whenever $(1-z) = 1/p$, with $p$ an integer, logarithmic terms will arise, corresponding
to conformal anomalies.

\bigskip

Next one can proceed to renormalize the on-shell action for $0 < z < 1/2$ as follows.
One substitutes these expansions into the regulated on-shell action:
\be
S = \frac{1}{2 \kappa_{d+1}^2} \int_{\cal M} d^{d+1} x \sqrt{-g} \left [ R + \Lambda
- \frac{1}{4} F_{mn} F^{mn} - \frac{1}{2} m^2 B_{m} B^{m} \right ] - \frac{1}{\kappa_{d+1}^2} \int_{\partial \cal M} d^d x \sqrt{-h} K
\ee
with the boundary regulated at $\rho = \ep$ and we now let $\kappa^2 \equiv \kappa_{3}^2$ in the case of interest, $d=2$.
The Gibbons-Hawking boundary term is included to ensure that the Dirichlet variational problem
is well-defined on the surface of fixed radius; note that $K$ is the trace of the second fundamental form.
This procedure results in a regulated action of the form:
\be
S_{\rm{reg}} = \frac{1}{2 \kappa^2} \int_{\rho = \ep} d^2{x} \sqrt{-g_{(0)}} \left [ \ep^{-1} a_{(0)} + \ep^{-z} a_{2 (1-z)} + \td{a}_2 \log \ep +
{\cal O} (\ep^0) \right ]
\ee
which involves a finite number of terms that diverge as $\ep \rightarrow 0$. Here
all coefficients $(a_{(k)}, \td{a})$ of divergent terms are local functions of the sources $(g_{(0) ab} (x),b_{(- z)a} (x))$:
\begin{eqnarray}
a_{(0)} &=& 2 \qquad a_{2 (1-z)} = -\frac{z}{2} b_{(-z)}g_{(0)}^{-1}b_{(-z)}, \\
\tilde{a}_2 &=& \text{Tr} g_{(2)} - z^{2} b_{(-z)}g_{(0)}^{-1} b_{(z)} = -\half R_{(0)}. \nonumber
\end{eqnarray}
These divergences can be removed using the following covariant counterterm action:
\be
S_{\rm ct} =  \frac{1}{2 \kappa^2} \int d^{2}x \sqrt{-\gamma} \left(- 2 + {z \over 2} B^a B_a + \hf R[\g] \log{\ep} \right),
\ee
where $\gamma$ is the induced metric. From the renormalized action, $S_{{\rm ren}} = S + S_{\rm ct}$, one can define the following
renormalized one point functions:
\be
\< V_{a} \> = - {1 \over \sqrt{-g_{(0)}}} {\delta S_{ren} \over \delta b_{(-z)}^{a}} = \lim_{\ep \to 0} \Big[ {\ep^{-z/2}
\over \sqrt{-\g}} {\delta S_{ren} \over \delta B^{a}} \Big] = -{z \over \kappa^{2}} b_{(z)a}, \label{vev1}
\ee
and for the stress energy tensor:
\bea
\< T_{ab} \> &=& - \frac{2}{\sqrt{-g_{(0)}}} \frac{\delta S_{\text{ren}}}{\delta g_{(0)}^{ab}} =
- \lim_{\ep \ra 0} \Big[ \frac{2}{\sqrt{-\g}} \frac{\delta S_{\text{ren}}}{\delta \g^{ab}} \Big]; \label{vev2} \\
&=&
{1 \over \kappa^{2}}\, \left[\,  g_{(2)ab} + \frac{1}{2} R(g_{(0)}) g_{(0)ab}
- {z \over 2} \left( b_{(-z)a}b_{(z)b} + b_{(z)a}b_{(-z)b} \right) - z (z-  \frac{1}{2}) \Big( b_{(-z) c} b_{(z)}^{c}  \Big) g_{(0)ab} \right]. \nn
\eea
Note that the answer for pure Einstein gravity, i.e. when $B_{m}  = 0$, agrees with that given in \cite{deHaro:2000xn}. Since the generating functional
of the dual field theory $W$ in Lorentzian signature is related to the renormalized on-shell action as $W = i S_{ren}$, these definitions for
the operators agree with those given in section \ref{field-ward}, as do
the dilatation and diffeomorphism Ward identities, which are respectively:
\bea
\< T_{a}^{a} \> &=& - {1 \over \kappa^{2}} \text{Tr}(g_{(2)}) = \frac{1}{\kappa^2} R_{(0)}
+ z b_{(-z)}^a \< V_a \> ; \\
D^{b} \< T_{ab} \> &=& \left( b_{(-z)a}\, D^{b} \< V_{b} \>  + F_{(-z)a}^{b} \< V_{b} \> \right). \nn
\eea
The relation between the bulk Newton constant $G_3$ and the central charge $c$ of the dual two-dimensional CFT is
\be
\frac{1}{\kappa^2} = \frac{1}{8 \pi G_3} = \frac{c}{24 \pi},
\ee
as derived in \cite {Henningson:1998gx}.

\subsection{Hamiltonian analysis}

In the previous section we showed that an increasing number of counterterms are needed
for $z > 1/2$. The renormalized one point functions and counterterms are in
such cases more conveniently computed using the Hamiltonian formulation of holographic renormalization. In this
section we will analyze holographic renormalization using the methods developed in
\cite{Papadimitriou:2004ap,Papadimitriou:2005ii}. These will allow us to compute renormalized correlation functions
for generic values of $z > 1/2$.

We begin by expressing the metric as
\be
ds^2 = g_{mn} dx^m dx^n = (N^2 + N_{a} N^a) dr^2 + 2 N_a dx^a dr + \gamma_{ab} dx^a dx^b,
\ee
where $N$ is the lapse and $N_a$ is the shift. The choices of $N=1$ and $N^a = 0$ make $r$ a Gaussian normal coordinate, related to the Fefferman-Graham
coordinate $\rho$ as $\rho = e^{-r}$.
In order to provide a Hamiltonian description of the dynamics one first expresses the curvature part of the action
in terms of quantities on hypersurfaces $\Sigma_r$, of constant $r$:
\be
S = \frac{1}{2 \kappa^2} \int d^3 x \sqrt{\gamma} N \left [ \hat{R} + K^2 - K_{ab} K^{ab} + \Lambda - \frac{1}{4} F_{mn} F^{mn} - \frac{1}{2} m^2 B_{m} B^m \right ],
\label{ham-act}
\ee
where $\hat{R}$ is the Ricci scalar of $\Sigma_r$ and $K_{ab}$ is its second fundamental form. After using the gauge freedom to fix
$N=1$ and $N_a =0$ the Einstein equations of motion become
\begin{eqnarray}
K^{2} - K_{a}^{b} K_{b}^{a} &=& \hat{R} + 2 \kappa^{2} T_{rr} \nn \\
D_{a} K_{b}^{a} - D_{b} K &=& \kappa^{2} T_{br} \\
\dot{K}_{a}^{b} + K K_{a}^{b} &=& \hat{R}_{a}^{b} - \kappa^{2} (T_{a}^{b} - \delta_{a}^{b} T) \nn
\end{eqnarray}
where $\dot{a}$ denotes $\pa_{r} a$ and
\be
\kappa^{2} T_{mn} = ( 1 - {1 \over 8} F^{2} - {1 \over 4} m^{2} B^{2} ) g_{mn} + \frac{1}{2} F_{mp} {F_{n}}^{p} + \frac{1}{2}
z^{2} B_{m} B_{n},
\ee
with $T = T^{m}_{m}$. Note that the $(ra)$ and $(rr)$ Einstein equations are the momentum and Hamilton constraints, which enforce that the
momenta conjugate to the lapse and shift functions vanish identically. The momentum conjugate to $B_r$ also vanishes (corresponding
to the divergence equation for the vector field) and the non-trivial canonical momenta are
\bea
\pi_{ab} &=& \pi_{ab}[\g , B_{c}]
= {\delta L \over \delta \dot{\g}^{ab}}
= {\delta I_{r} \over \delta \g^{ab}}
= -{1 \over 2 \kappa^{2}} \sqrt{\g} (K\g_{ab} - K_{ab}), \label{momenta-ab} \\
\pi_{a} &=& \pi_{a}[\g , B_{b}] = {\delta L \over \delta \dot{B}^{a}}
= {\delta I_{r} \over \delta B^{a}} = -{1 \over 2 \kappa^{2}}  \sqrt{\g} F_{ra}
= -{1 \over 2 \kappa^{2}}  \sqrt{\g} ( \dot{B_{a}} - \partial_{a}B_{r} )., \label{momenta-a}
\eea
where $I_{r} = \big[ \int dr L \big]_{\rm on-shell}$ is the on-shell action.
This implies that the extrinsic curvature $K_{ab}$ and the momenta of the vector field $B_a$ are
themselves functionals of the induced fields on $\Sigma_r$.
Note that the extrinsic curvature is given by
\be
K_{ab} = \frac{1}{2} n^{m} \partial_{m} \g_{ab} = \frac{1}{2} \dot{\g}_{ab},
\ee
where $n$ is the normal to $\Sigma_r$.
In the Hamiltonian version of holographic renormalization one uses the equations of motion to determine the asymptotic form of the momenta
as functionals of the induced fields. This method has the key advantage of maintaining covariance at all stages, thus ensuring that Ward identies are manifest
and it also shortens the computation of counterterms.

In the method of holographic renormalization used in the last section
the asymptotic analysis begins by expanding the bulk fields in the $\rho$ coordinate. In the Hamiltonian method one notes that
the non-normalizable modes of the induced fields behave asymptotically as
\begin{eqnarray}
&& \g_{ab} \sim e^{2r} g_{(0)ab}, \qquad  \dot{\g}_{ab} \sim 2 \gamma_{ab}, \\
&& B_{a} \sim e^{zr} b_{(-z)a}, \qquad \dot{B}_{a} \sim z B_{a}. \nn
\end{eqnarray}
Note that the field $B_r$ is entirely determined by these fields, using the vector divergence equation.
The dilatation operator, identified with the functional form of the asymptotic $r$-derivative in the solution space, is found to be:
\be
\partial_{r}
= \int d^{2}x \left ( \dot{\g}_{ab} {\delta \over \delta \g_{ab}} + \dot{B}_{a} {\delta \over \delta B_{a}} \right )
\sim \int d^{2}x \left ( 2 \g_{ab} {\delta \over \delta \g_{ab}} + z B_{a} {\delta \over \delta B_{a}} \right ) \equiv \delta_{D}. \label{dil}
\ee
Since $K_{ab}, \dot{B}_{a}$
and $B_{r}$ are functionals of the induced fields, each can be written asymptotically as an expansion in eigenfunctions
of the dilatation operator, \eqref{dil}.
Furthermore, the leading terms in the asymptotic radial expansions coincide with those in the asymptotic expansions in eigenfunctions of the dilatation operator.
This allows one to write:
\bea
{K_{a}}^{b} &=&
{K_{(0)a}}^{b} + {K_{(\a_{1})a}}^{b} + {K_{(\a_{2})a}}^{b} + \cdots
+ {K_{(2)a}}^{b} + {\ \tilde{K}_{(2)a}}^{b} \log{e^{-2r}} + \cdots, \qquad {K_{(0)a}}^{b} = \delta_{a}^{b} \nn \\
\dot{B}_{a} &=& \dot{B}_{(-z)a} + \dot{B}_{(\b_{1})a} + \dot{B}_{(\b_{2})a} + \cdots, \qquad \dot{B}_{(-z)a} = z B_{a} \label{ham-p} \\
B_{r} &=& B_{(2 - z)r} + B_{(\s_{1})r} + \cdots, \nn
\end{eqnarray}
where the dilatation weights are such that
\begin{eqnarray}
\delta_{D} {K_{(n)a}}^{b} &=& - n {K_{(n)a}}^{b}, \quad n < 2 \\
\delta_D \dot{B}_{(n) a} &=& - n \dot{B}_{(n)a}, \qquad
\delta_D B_{(n)r} = - n B_{(n)r}. \nn
\end{eqnarray}
with the logarithmic terms similarly transforming homogeneously.
Note that $\delta_D K_{(n)ab} = - (n - 2) K_{(n)ab}$ and $[\delta_D, \partial_{a}] = 0$ but $[\delta_D, \partial_{r}] \neq 0$.
The term ${K_{(2)a}}^{b}$ transforms as
\be
\delta_D K_{(2)a}^{b} = - 2 K_{(2)a}^{b} - 2 \tilde{K}_{(2)a}^{b}.
\ee
This inhomogeneous transformation is obtained
by requiring firstly that $\delta_D$ does not act on coordinates (i.e, on the logarithm)
and secondly that the action of $\partial_{r}$ on $K_{a}^{b}$
provides asymptotically the same result as the action of $\delta_D$, where $\partial_{r} {K_{(d)a}}^{b} \sim - d {K_{(d)a}}^{b}$.
Using the vector field equations and divergence equation, one can show that
\be
D_{a}B^{a} = - {1 \over \sqrt{\g}} \partial_{r} ( \sqrt{\g}B_{r} ), \qquad
\dot{B}_{r} = z B_{(2-z)r} - K B_{r},
\ee
and hence the expansion for $\dot{B}_{r}$ can indeed be written in terms of the expansion for $K$ and $B_{r}$.

The expansions of the momenta in eigenfunctions of the dilatation operator can be determined iteratively by solving the field equations.
One can now deduce immediately the first subleading term ${K_{(\a_{1})a}}^{b}$ of $K_{a}^{b}$ by looking at the leading order terms
in the Einstein equations. The $(ra)$ equation implies that
\be
D_{b}K_{a}^{b} - D_{a} K = \kappa^{2} T_{a r}
= \frac{1}{2} F_{ab} \g^{cb} ( \dot{B}_{c} - \partial_{c} B_{r} ) + \frac{1}{2} z^{2} B_{a} B_{r}.
\ee
Since $D_{b}{K_{(0)a}}^{b} - \partial_{a} K_{(0)} = 0$, the lowest order terms contributing are
\be
D_{b}{K_{(\a_{1})a}}^{b} - \partial_{a} K_{(\a_{1})}
= {z \over 2} F_{ab} \g^{cb}B_{c} + {z^{2} \over 2} B_{a} B_{(2-z)r}.
\ee
This implies that $\a_{1} = 2(1-z)$. One can then use this fact in the $(ab)$ Einstein equations to find ${K_{(\a_{1})a}}^{b}$,
resulting in
\be
{K_{(2-2z)a}}^{b} = - {z \over 2} \big( B_{a} B^{b} - \hf ( B\g^{-1}B ) \delta_{a}^{b} \big). \label{k2-2z}
\ee
Note that $K_{(2 - 2 z)} := {\ K_{(2-2z)i}}^{i} = 0$. One can derive similar equations for further coefficients in \eqref{ham-p} but the
ordering of the weights $(\alpha_{(n)}, \beta_{(n)})$ depends on the value of $z$. For example, when $z < 1/2$, the coefficient $\beta_1 = z$ is
the first subleading term in the vector field expansion, as we showed in the previous section, whilst for $z > 1/2$, the first subleading term
is instead $B_{ (2- 3z)a}$ since $(2 - 3 z) < z$. At $z = 1/2$ one needs to include logarithmic terms, related to the conformal anomalies,
to satisfy the field equations.

Before solving for further coefficients,
let us discuss how this information will be used to determine the renormalized on-shell action and one point functions. Starting
from \eqref{ham-act} one can differentiate the on-shell action with respect to $r$ to obtain
\be
\dot{S}_{\rm on-shell} = \frac{1}{\kappa^2} \int_{\Sigma_r} d^2x \sqrt{\gamma} \left ( \hat{R} + 1 - \frac{1}{4} F_{ab} F^{ab} - \frac{1}{2} z^2 B_a B^a \right ).
\ee
One can then write the regulated action as
\be
I_r = {1 \over \kappa^{2}} \int_{\S_{r}} d^{d}x \sqrt{\g} ( K - \l ),
\ee
where $\l$ satisfies
\be
\dot{\l} + K \l - \kappa^{2} (2 + \frac{1}{4} F^2 + \frac{1}{2} z^2 B^2) = 0.
\ee
The variable $\lambda$ admits an expansion in dilatation eigenfunctions:
\be
\lambda = \l_{(\e_{0})} + \l_{(\e_{1})} + ... + \l_{(2)} + \tilde{\l}_{(2)} \log{e^{-2r}} + \cdots
\ee
where each term transforms homogeneously, namely $\delta_D \l_{(n)} = - n \l_{(n)}$
except for $\lambda_{(2)}$.  The transformation law for the latter is obtained in a similar fashion as that for ${K_{(2)a}}^{b}$.
Terms in the on-shell action are divergent as $r \rightarrow \infty$ only for $n < 2$, along with the logarithmic term, and thus the
counterterm action is formally given by
\be
I_{\rm ct} = - {1 \over \kappa^{2}} \int_{\S_{r}} d^{2}x \sqrt{\g} \left ( \sum_{n < 2} ( K_{(n)} - \l_{(n)} ) - \tilde{\l}_{(2)} \log{e^{-2r}} \right ).
\label{ct-action}
\ee
The terms in the dilatation expansion of $\l$ can be
obtained by iteratively solving the above first order equation defining $\l$, but a more efficient procedure is the following. Note first that
\begin{eqnarray}
2 \g_{ab} \pi^{ab} + z B_{a} \pi^{a}
&=& 2 \g_{ab} {\delta I_{r} \over \delta \g_{ab}} + z B_{a} {\delta I_{r} \over \delta B_{a}} \\
&=& {1 \over \kappa^{2}} \int_{\S_{r}} d^{2}x \sqrt{\gamma}
\left [ 2 \g_{ab} {\delta \over \delta \g_{ab}} + z B_{a} {\delta \over \delta B_{a}} \right ] ( K - \l ). \nn
\end{eqnarray}
Hence,
\begin{equation}
2 \g_{ab} \pi^{ab} + z B_{a} \pi^{a}
= {1 \over \kappa^{2}} \delta_D ( \sqrt{\g} ( K - \l ) ) \Lra ( 1 + \delta_D ) K
+ {z \over 2} B_{a} \g^{ab} ( \dot{B}_{b} - \partial_{b} B_{r} ) = ( 2 + \delta_D ) \l\, \label{ct-def}
\end{equation}
where one has used (\ref{momenta-ab}) and (\ref{momenta-a}), together with: $\delta_{D}\sqrt{\g} = 2\sqrt{\g}$ which follows from the definition of $\delta_{D}$.
This last equation then allows the iterative
determination of the expansion of $\l$. For example, looking at the leading order term one finds that
\be
K_{(0)} = ( 2 + \delta_D ) \l_{(0)} \qquad \rightarrow \qquad  \l_{(0)} = 1.
\ee
The first subleading term has weight $2 (1-z)$ and is given by
\be
\frac{z^2}{2} (B \gamma^{-1} B) = (2 + \delta_{D}) \lambda_{(2 - 2z)} \leftrightarrow  \lambda_{(2 -2 z)} = \frac{z}{4} (B \gamma^{-1} B).
\ee
As mentioned already above, the question of which terms appear at subsequent order depends on the value of $z$. For $z < 1/2$, the only other divergent term
is the logarithmic term, which follows from solving \eqref{ct-def} at weight two. Using the expression for $K_{(2)}$ given in \eqref{k2-ex}
one finds that
\be
(2 + \delta_{\cal D})(\lambda_{(2)} + \td{\lambda}_{(2)} {\rm log} e^{-2r}) = - 2 \td{\lambda}_{(2)} = - \frac{1}{2} R.
\ee
For $z < 1/2$ this suffices to determine explicitly the counterterm action
\be
I_{\rm ct} = {1 \over 2\kappa^{2}} \int_{\S_{r}} d^{2}x \sqrt{\g} \big( - 2 + {z \over 2}(B\g^{-1}B) + \hf R \log{e^{-2r}} \big),
\ee
in agreement with that found in the previous section. Further counterterms are needed for $z \ge 1/2$ but, as we will see, the explicit form
is not needed to compute renormalized correlation functions for non-rational values of $z$.

In general, the renormalized on-shell action is given by
\be
I_{\text{ren}} = \lim_{r \rightarrow \infty} ( I_{r} + I_{\rm ct} ) = {1 \over \kappa^{2}} \int d^{2} x \sqrt{\g}  ( K_{(2)} - \l_{(2)})\ . \label{ren-act}
\ee
The one-point functions can be determined by using the Hamilton-Jacobi relations, which can be written as:
\be
\pi^{ab} \delta \g_{ab} + \pi^{a} \delta B_{a} = {1 \over \kappa^{2}} \int_{\S_{r}} d^{2} x \delta [ \sqrt{\g}  ( K - \l)].
\ee
Taking $r \rightarrow \infty$,
one expands the momenta and the integrand in eigenfunctions of the dilatation operator and matches terms with the same weight.
The procedure implies in particular that:
\be
\delta I_{\text{ren}} = {1 \over \kappa^{2}} \int d^{2} x \delta \left [ \sqrt{\g}
( K_{(2)} - \l_{(2)}) \right ] = \left [ \pi_{ab} \g^{ac}\g^{be} \delta \g_{ce} \right ]_{(0)} + \left [ \pi_{a}\, \g^{ab} \delta B_{b} \right ]_{(0)}\, ,
\ee
where the subscript represents the overall terms with zero dilatation weight. Since, by the definition of $\delta_{D}$, the vector field has
weight $z$, the induced metric weight 2 and its inverse weight -2, the renormalized one-point functions are then found to be:
\begin{eqnarray}
\< T_{ab} \> &=&
- \frac{2}{-\sqrt{g_{(0)}}} \frac{\delta I_{ren}}{\delta g_{(0)}^{ab}} = \lim_{r \to \infty}
\Big[ \frac{2}{-\sqrt{\g}} \frac{\delta I_{\text{ren}}}{\delta \g^{ab}} \Big]
= {1 \over \kappa^{2}} \Big[ K_{(2)} \g_{ab} - {K_{(2)a}}^{c}\g_{cb} \Big]; \label{ham-res} \\
\< V_{a} \> &=&  - {1 \over \sqrt{-g_{(0)}}} {\delta I_{ren} \over \delta b_{(-z)}^{a}}
= \lim_{r \to \infty} \Big[ {e^{zr} \over \sqrt{-\g}} {\delta I_{ren} \over \delta B^{a}} \Big]
= {1 \over 2 \kappa^{2}} \lim_{r \to \infty} \Big[ e^{zr} \dot{B}_{(z)a} \Big]. \nn
\end{eqnarray}
It should be emphasized that these expressions for the renormalized one point functions hold for general values of $z < 1$, as does
the form \eqref{ren-act} for the renormalized action. However, one still needs to determine the relation between the momenta coefficients
and coefficients in the asymptotic expansions of the fields, which in general can involve both the normalizable modes and local functionals
of the sources.

When $z \neq (1 - \frac{1}{n})$, with $n$ an integer, the map between momenta coefficients and terms in the asymptotic expansions is particularly simple. Let us
express the asymptotic expansions as in the previous section as
\bea
\gamma_{ab} &=& g_{(0) ab} + \cdots + e^{-2r} g_{(2) ab} + \cdots \\
B_{a} &=& e^{z r} ( b_{(-z) a} + \cdots) + e^{-z r} (b_{(z) a} + \cdots), \nn
\eea
where $\rho = e^{-2r}$. Then,
\bea
K_{(2)} &=& \frac{1}{2} \left ( R[g_{(0)}] - 2 z^2 b_{(-z)}^a b_{(z)a} \right ); \label{k2-ex} \\
K_{(2) ab} &=& - g_{(2) ab} + \frac{z}{2} (b_{(-z) a} b_{(z) b} + b_{(z)a} b_{(-z) b} ) - \frac{z}{2} b_{(-z) c} b_{(z)}^c g_{(0) ab}; \nn \\
\Big[ e^{zr} \dot{B}_{(z)a} \Big] &=& - 2 z b_{(z)}. \nn
\eea
and substituting into the renormalized one point functions \eqref{ham-res} results in the same expressions as \eqref{vev1} and \eqref{vev2}.

When $z = (1 - \frac{1}{n})$, with $n$ an integer, functionals of the vector operator source can have the required dilatation weight to contribute
to the one point functions. In such cases there are additional contributions to the map between momenta coefficients and terms in the asymptotic expansions,
and one has to compute the one point functions on a case-by-case basis. For example, in the case of $z= 1/2$
\be
K_{(2)} = \frac{1}{2}  \left ( R[g_{(0)}] - 2 z^2 b_{(-z)}^a b_{(z)a} \right ) + \frac{1}{2} K_{(2-2z) a}^{\;b} K_{(2-2z)b}^{a},
\ee
where, using \eqref{k2-2z},
\be
K_{(2-2z) a}^{\;b} = - \frac{z}{2} \left ( b_{(-z)a} b_{(-z)}^b - \frac{1}{2} (b_{(-z)c} b_{(-z)}^c) \delta^{b}_a \right ).
\ee
This implies that the conformal anomaly is given by
\be
\< T^{a}_{a} \> = \frac{1}{\kappa^2} \left ( R_{(0)}
- z^2 b_{(-z)}^a b_{(z) a} + \frac{z^2}{16} (b^{a}_{(-z)} b_{(-z) a})^2 \right ),
\ee
and thus involves a local functional of the vector field source.

\subsection{Analysis for $z > 1$}

Let us now discuss the issues that arise when $z > 1$ and the vector field is dual to an irrelevant operator in the conformal field theory.
Since irrelevant operators modify the UV behavior of the quantum field theory, their sources can only be treated perturbatively, which
allows their correlation functions to be computed. The holographic analogue can be seen in \eqref{exp1}: even for $z > 1$
the data $(g_{(0) ab}, g_{(2) ab}, b_{(-z)a}, b_{(z)a})$ supplies the independent integration constants for the bulk equations, but when
$z> 1$ the limit of $g_{ab} (\rho \rightarrow 0)$ is no longer finite. In fact, using \eqref{exp2}, one sees that the metric
\be
g_{ab} = \sum_{m,n} g_{2 m + 2 n (1-z)} \rho^{m + n (1-z)}
\ee
contains terms for $m=0$ and $n \ge 0$ which behave as
\be
g_{- 2 n(z-1)} \rho^{-n (z-1)} \sim b_{(-z)}^{2n} \rho^{-n (z-1)},
\ee
and thus terms which are higher order in the vector operator source diverge faster as $\rho \rightarrow 0$, as expected. Working at finite $b_{(-z)}$
an infinite number of counterterms would thus in general be needed. A well-defined
problem is obtained by working perturbatively with small  $b_{(-z)a}$ such that
\be
| b_{(-z)} |^2 \ll \ep^{z-1},
\ee
where $\rho = \ep$ is the cutoff. To compute an $n$-point function of the dual vector operator, one should only retain terms to
order $b_{(-z)}^n$ and thus only a finite number of counterterms are needed. Logarithmic terms in the on-shell action related to conformal
anomalies can arise whenever
\be
z = 1 + \frac{p}{q},
\ee
where $(p,q)$ are integers. Except in such cases, where $z$ is rational, the renormalized one point functions are just as for $z < 1$, i.e.
\bea
\< V_{a} \>  &=& -{z \over \kappa^{2}} b_{(z)a}; \\
\< T_{ab} \> &=& {1 \over \kappa^{2}}\, \left[\,  g_{(2)ab} + \frac{1}{2} R(g_{(0)}) g_{(0)ab}
- {z \over 2} \left( b_{(-z)a}b_{(z)b} + b_{(z)a}b_{(-z)b} \right) - z (z-  \frac{1}{2}) \Big( b_{(-z) c} b_{(z)}^{c}  \Big) g_{(0)ab} \right]. \nn
\eea
To prove this, one can use the Hamiltonian method of the previous section: provided that the source is treated perturbatively, the dilatation operator
is well-defined and the momenta admit expansions in eigenfunctions of this dilatation operator. The general expressions for the renormalized one-point
functions in terms of the momenta coefficients given in \eqref{ham-res} can then immediately
be rewritten in terms of coefficients in the asymptotic expansion when $z$ is not rational, as terms involving only the vector field sources cannot have the correct
dilatation weight. For rational values of $z$ the map between the momenta and asymptotic coefficients can indeed involve polynomials in the vector
field sources, and it needs to be worked out iteratively on a case by case basis.

Note that in the Hamiltonian method one does not actually need to explicitly compute the counterterms $\lambda_{(n)}$ to derive the correlation functions,
although they would be needed to compute the on-shell value of the action. Formally, at least, one can work to arbitrarily high perturbative order in the operator
source $b_{(-z)a}$, with corresponding counterterms of increasing order of divergence. If however the source $b_{(-z) a}$ is treated as finite, then there is no
well-defined asymptotic, or equivalently dilatation, expansion and the counterterm action \eqref{ct-action} is not a priori valid. This corresponds to the
fact that switching on a generic finite deformation by the dual vector operator makes the dual quantum field theory non-renormalizable.

In the case of interest here, however, the source $b_{(-z)a}$ is finite but null: just as in the field theory discussion earlier,
we can compute correlators of the vector operator in the deformed theory by setting
\be
g_{(0) ab} = \eta_{ab}; \qquad b_{(-z) a} \equiv b \delta_{au} + a_{(-z)a}, \label{back}
\ee
where the source $a_{(-z)a}$ is treated perturbatively. The existence of a dilatation symmetry is preserved at finite $b$ and all bulk
fields still admit an asymptotic expansion in terms of eigenfunctions of the dilatation operator, even though the metric $g_{ab}$ does not have a finite
limit as $\rho \rightarrow 0$.

Now consider the following: treating $b_{(-z)a}$ perturbatively first derive the counterterm action \eqref{ct-action}, working
recursively in powers of the source. Then
\be
I_{\rm ct} = - {1 \over \kappa^{2}} \int_{\S_{r}} d^{2}x \sqrt{\g} \left ( \sum_{n < 2} ( K_{(n)} - \l_{(n)} ) - \tilde{\l}_{(2)} \log{e^{-2r}} \right ),
\ee
where in addition to the counterterms $\l_{(0)}$, $\td{\l}_{(2)}$ and $\lambda_{- 2 (z-1)}$ computed explicitly earlier there are an infinite
number of counterterms at $z > 1$. For example, polynomials of the vector field occur,
\be
\lambda_{- 2n (z-1)} = c_{2 n (z - 1)} (B^a B_a)^{n}
\ee
where the coefficients $c_{2n (z-1)}$ may be determined iteratively in $n$, working perturbatively in the source. This counterterm is the holographic
analogue of \eqref{ct1} and counterterms involving further derivatives and curvatures will also occur.
If these counterterms are evaluated
on the anisotropic background itself \eqref{back} in which $a_{(-z) a} = 0$, then, since the source is both null and constant,
all counterterms apart from $\l_{(0)}$ vanish. This is the holographic analogue of the deformation being exactly marginal with respect to
the anisotropic symmetry. To compute the two point function of the vector operator in the deformed theory we will need to retain terms in the
action to order $a_{(-z)}^2$, and following the arguments of section \eqref{counter} there will be a finite number of terms for $z < 2$.

\section{Linearized analysis around chiral background} \label{se:linear}

In this section we will consider the linearized equations of motion around the chiral background for generic values of $z$ in two dimensions, and the corresponding
two point functions of the stress energy tensor and vector operator in the deformed theory. We should note that the analysis excludes those values of $z$
for which the deforming operator itself acquires an anomalous dimension; the case of $z=2$, Schr\"{o}dinger, is one such example, which was analyzed in detail
in \cite{Guica:2010sw}.

\subsection{Linearized equations}

Let us perturb the fields around the background as:
\begin{eqnarray}
B_{m} &=& \rho^{-z/2} b_{m}(x,\rho) = b\, \rho^{-z/2} \delta_{m}^{u} + {a}_{m}(\rho,u,v), \\
ds^2 &=& \frac{d \rho^2}{4 \rho^2} + \frac{1}{\rho} g_{ab} (x,\rho) dx^a dx^b, \qquad
g_{ab} = \bar{h}_{ab}(\rho) + h_{ab}(\rho,u,v), \nonumber
\end{eqnarray}
where
\be
\bar{h}_{ab} dx^a dx^b \equiv (2 du dv +  \sigma^2 \rho^{1-z} du^2).
\ee
The linearized Einstein equations can then be written as:
\begin{eqnarray}
&& R_{ab}[h] + \tr(\bar{h}^{-1} h') \bar{h}_{ab} - \rho \left( 2\, h_{ab}'' - 2\, z\, b^{2} \delta_{(a}^{u}h_{b)v}'\, \rho^{-z} +
{z \over 2}\, b^{2}\, \tr( \bar{h}^{-1} h' ) \delta_{a}^{u} \delta_{b}^{u}\, \rho^{-z} \right) \label{spatial} \\
& &
= \hf ( 1 + z ) z\, b^{2} h_{vv} \bar{h}_{ab}\, \rho^{-z} + \big( {z \over 2}\, b^{2} \big)^{2} h_{vv} \delta_{a}^{u}
\delta_{b}^{u}\, \rho^{1-2z} + z^{2}\, b\, {a}_{(a} \delta_{b)}^{u}\, \rho^{-z/2} \nonumber \\
&& \qquad + 2\, z\, b\, \left( \delta_{(a}^{u} f_{b)\rho}
- f_{v\rho}\, \bar{h}_{ab} \right) \rho^{1-z/2};\nonumber \\
& & \partial_{a} \left( \tr( \bar{h}^{-1}h') \right) - \bar{h}^{bc} \partial_{c}h_{ab}' -
{1 \over 4} z\, b^{2} \partial_{a} h_{vv}\, \rho^{-z} + \hf z\, b^{2} \delta_{a}^{u}\, \rho^{-z}\, \partial_{c} \left( h_{v}^{c}
- \hf \tr(h) \delta_{v}^{c} \right) \label{cross} \\
& & \qquad = \hf z\, b\, f_{av}\, \rho^{-z/2} - z^{2} b\, {a}_{\rho}\, \delta_{a}^{u}\, \rho^{-z/2}\ ; \nonumber \\
& & {1 \over 4} z\, b^{2}\, \partial_{\rho} \left( \rho^{-z}\, h_{vv} \right) = \hf \tr( \bar{h}^{-1} h'' ) \label{radial} \\
& & R_{ab}[h] - \hf \bar{h}_{ab} R[h] = 0. \label{Ricci}
\end{eqnarray}
The last equation is the linearization of the two-dimensional identity $R_{ab}[g] - \hf g_{ab}R[g] = 0$. Note also that
we define $f_{ab} := \partial_{a} {a}_{b} - \partial_{b} {a}_{a}$ as the curvature of the vector fluctuation $a_{b}$.

The linearized vector field equations are
\begin{eqnarray}
\partial_{a} \left( \bar{h}^{ab} f_{b\rho} \right) - {z \over 2}\, b\, \rho^{-1-z/2}  \partial_{a} \left( h_{v}^{a} - \hf\, \tr( h )
\delta_{v}^{a} \right) &=& {z^{2} \over \rho}\, {a}_{\rho}; \label{lin-rho-eq1} \\
\partial_{a} \left( \bar{h}^{ab} \bar{h}^{cd} f_{bd} \right) + 4 \partial_{\rho} \left( \rho \bar{h}^{ac}
f_{\rho a} \right) + 2\, z\, b\, \rho^{-z/2} \partial_{\rho} \left( h_{v}^{c} - \hf\, \tr( h ) \delta_{v}^{c} \right) &=& {z^{2} \over \rho}\,
\bar{h}^{ca} a_{a}. \label{lin-a-eq1}
\end{eqnarray}
whilst the linearized divergence equation is:
\be
\partial_{a} \left( \bar{h}^{ac} a_{c} \right) + 4 \rho\, a_{\rho}' - b\, \rho^{-z/2} \partial_{a} \left( h_{v}^{a} - \hf\, \tr( h )
\delta_{v}^{a} \right) = 0. \label{lin-div-eq1}
\ee
It is also useful to write:
\be
\tr(h) = -\sigma^{2} h_{vv} \rho^{1-z} + 2 h_{uv}; \qquad \partial_{a} \left( h^{a}_{v} - \frac{1}{2} \tr(h) \delta_{v}^{a} \right)
= - \frac{1}{2} \sigma^{2} h_{vv,v} \rho^{1-z} + h_{vv,u}.
\ee
One now begins with the identity (\ref{Ricci}), which implies that
$R_{vv} = 0\, ,\ R_{uu} = \s^{2} \rho^{1-z} R_{uv}$ , where the components of the Ricci tensor are:
\begin{eqnarray}
R_{uv}[h] &=&  \hf h_{vv,uu} + \hf h_{uu,vv} - h_{uv,uv}; \\
R_{uu}[h] &=&  \s^{2} \rho^{1-z}\, \left( \hf h_{vv,uu} + \hf h_{uu,vv} - h_{uv,uv} \right). \nonumber
\end{eqnarray}
Using these identities, the ${(vv)}$ component of the Einstein equations is solved by
\be
h_{vv}'' = 0 \qquad \rightarrow \qquad h_{vv} = h_{(0) vv} + \rho h_{(2)vv},
\ee
where both $h_{(0)vv}$ and $h_{(2)vv}$ are arbitrary functions of $(u,v)$.
The $(v)$ component of (\ref{cross}) together with (\ref{radial}) lead to:
\be
h_{uv,v}' - h_{vv,u}' - \qt\, z\, b^{2}\, h_{vv,v}\, \rho^{-z} = 0; \qquad
h_{uv}'' = \qt\, z\, b^{2}\, \partial_{\rho} \left( \rho^{-z}\, h_{vv} \right).
\ee
Integrating the second of these equations gives:
\be
h_{uv} = h_{(0)uv} + \rho h_{(2) uv} + \frac{b^2 z}{4} \left ( \frac{1}{(1-z)} \rho^{1-z} h_{(0) vv} + \frac{1}{(2-z)}
\rho^{2-z} h_{(2) vv} \right ),
\ee
whilst the first equation implies that:
\be
\pa_v h_{(2) uv} = \pa_u h_{(2) vv}.
\ee
The other Einstein equations do not decouple from the vector field fluctuations. One can however use equation \eqref{Ricci}
to express the remaining graviton fluctuation as
\bea
h_{uu} &=& \td{H}_{(0) uu} + \rho \tilde{H}_{(2) uu} - \frac{z}{4 (1-2z)} \sigma^4 h_{(0) vv} \rho^{2 - 2 z} \nn \\
&& - \frac{z}{4 (3 - 2z)} \sigma^4 \rho^{3-2z} h_{(2)vv} + \frac{\sigma^2}{(2-z)} \rho^{2-z} h_{(2) uv} + h^{V}_{uu},
\eea
where $(\td{H}_{(0) uu}, \tilde{H}_{(2)uu})$ are integration constants and $h^{V}_{uu}$ is defined as the solution to
\be
\pa_{\rho}^2 h^{V}_{uu} = - \frac{1}{2} z b \sigma^2 \rho^{1-z} \pa_{\rho} (\rho^{-z/2} a_{v}) + z b \pa_{\rho} (\rho^{-z/2} a_{u}) + z b \rho^{-z/2}
( \frac{1}{2} \sigma^2 \rho^{1-z} a_{\rho,v} - a_{\rho,u}).
\ee
In order to solve the remaining Einstein equations, the graviton fluctuation must in addition
satisfy the $u$ component of \eqref{cross} and the $(uv)$ component of \eqref{spatial}, which
requires
\bea
h_{uu,v}' &=& \frac{1}{2} z b^{2} \rho^{-z} (h_{(0)vv,u} + \rho h_{(2) vv,u}) + \pa_u h_{(2) uv} + X; \\
X &=& \frac{1}{2} z b \rho^{-z/2} \left( a_{u,v} - a_{v,u} + 2 z a_{\rho} \right); \nn \\
h_{uu,vv} &=& 2 h_{uv,uv} - h_{vv,uu} - 4 h_{(2) uv} + 2 z \sigma^2 \rho^{1-z} h_{(2)vv} + Y; \\
Y &=& 2 z b \rho^{1-z} \left( \partial_{\rho} \left( \rho^{z/2}\, {a}_{v} \right) - \rho^{z/2}\, a_{\rho,v} \right). \nn
\eea
As we will show below,
these constraints impose a restriction on the integration constant $\tilde{h}_{(2)uu}$, related to the diffeomorphism Ward identity.
These equations are automatically satisfied when the vector field equations are solved.
We use the notation $(\td{H}_{(0) uu}, \td{H}_{(2) uu})$ to denote the integration constants anticipating the fact that
$h^{V}_{uu}$ could also contribute terms at order $\rho^0$ and $\rho$ in the asymptotic expansion as $\rho \rightarrow 0$.

The linearized vector field equations can be written in terms of the metric fluctuations as follows. The divergence equation
\eqref{lin-div-eq1} becomes
\be
4 \rho a_{\rho}' = - a_{v,u} - a_{u,v} + \s^{2} a_{v,v} \rho^{1-z} +
b \rho^{-z/2} \left( - \frac{1}{2} \s^{2} \partial_{v} (h_{(0) vv} + \rho h_{(2)vv})
\rho^{1-z} + \partial_{u} (h_{(0)vv} + \rho h_{(2) vv}) \right).  \label{lin-div-eq}
\ee
Equation \eqref{lin-rho-eq1} becomes
\bea
z^{2}  \rho^{-1} a_{\rho} &=&
\partial_{v} \left( a_{\rho,u} - a_{u}' \right) + \partial_{u} \left( a_{\rho,v} - a_{v}' \right) - \s^{2} \rho^{1-z} \partial_{v}
\left( a_{\rho,v} - a_{v}' \right)
\label{lin-rho-eq} \\
&&
 - \frac{1}{2} z b \rho^{-1-z/2}
\left( - \frac{1}{2} \s^{2} \partial_{v} (h_{(0) vv} + \rho h_{(2) vv})
\rho^{1-z} + \partial_{u} (h_{(0)vv} + \rho h_{(2) vv}) \right). \nn
\eea
Equations \eqref{lin-a-eq1} become
\bea
\rho^{z/2} \partial_{\rho}
\left[ \rho^{1- z}\, \partial_{\rho} \left( \rho^{z/2}\, a_{v} \right) \right]
&=& \partial_{\rho} \left( \rho\, a_{\rho,v} \right) - \frac{1}{4}
\partial_{v} \left( a_{v,u} - a_{u,v} \right) - \frac{1}{2} z b \rho^{-z/2} h_{(2) vv}; \label{lin-u-eq} \\
\rho^{z/2} \partial_{\rho} \left[ \rho^{1 - z}\, \partial_{\rho}
\left( \rho^{z/2}\, a_{u} \right) \right] &=&
 - \frac{1}{4} \partial_{u} \left( a_{u,v} - a_{v,u} \right)
+ \s^{2} \rho^{1-z} \rho^{z/2} \partial_{\rho} \left[ \rho^{1 - z} \partial_{\rho} \left( \rho^{z/2} a_{v} \right) \right] \nn \\
& & + \partial_{\rho} \left( \rho a_{\rho,u} \right)
- \s^{2} \partial_{\rho} \left( \rho^{2-z} a_{\rho,v} \right) + (1-z) \s^{2} \rho^{1-z} a_{v}' \label{lin-v-eq} \\
&& + \frac{1}{4} z(1-z) b \s^{2} \rho^{-3z/2} h_{(0)vv} + \frac{1}{4} z (2 - z) \s^2 \rho^{1 -3 z/2} h_{(2) vv}. \nn
\eea
These field equations can be diagonalized to give fourth order differential equations.
To show this, let us first define the differential operator
\be
\Delta := \rho \pa_{\rho}^2 + \pa_{\rho} - \frac{z^2}{4} \rho^{-1} + \frac{1}{2} \pa_{u} \pa_{v} - \frac{\sigma^2}{4} \rho^{1-z} \pa_{v}^2. \label{Delta}
\ee
We then define
\be
a_{v}^{V} \equiv a_v - \frac{b}{2} \rho^{-z/2} (h_{(0) vv} + \rho h_{(2) vv}),
\ee
as well as
\be
a^{V}_{\rho} \equiv a_{\rho} - \frac{1}{2} b \pa_{u}^{-1} h_{(2) uv} \rho^{-z/2},
\ee
where the inverse derivative is abbreviated notation such that
\be
A = \pa^{-1} B \qquad \rightarrow \qquad \pa A = B.
\ee
(In practice the solutions are expressed in momentum space, where the inverse derivative acts by
division of momenta.)
By differentiating \eqref{lin-div-eq} with respect to $v$ and inserting into \eqref{lin-u-eq} one obtains
\be
\Delta a_v^V = \pa_v a^V_{\rho}. \label{2nd1}
\ee
Differentiating \eqref{lin-div-eq} with respect to $\rho$ and subtracting it from \eqref{lin-rho-eq} one obtains
\be
\Delta a^V_{\rho} = \frac{\sigma^2}{4} (1-z) \rho^{-z} \pa_v a_v^V. \label{2nd2}
\ee
Combing these equations, one finds that $a^{V}_{\rho}$ satisfies the fourth order equation
\be
\rho^{z} \Delta (\rho^z \Delta a^{V}_{\rho}) = \frac{\sigma^2}{4} (1-z) \rho^z \pa_{v}^2 a^{V}_{\rho},
\ee
whilst $a_v^V$ also satisfies a fourth order equation
\be
\Delta^2 a_v^V = \frac{\sigma^2}{4} (1-z) \rho^{-z} \pa_v^2 a_v^V.
\ee
Given the solutions for $a_v^V$ and $a_{\rho}^V$ one can then determine $a_u$ using the remaining vector field equations; one
first determines $a_{u}$ using (\ref{lin-div-eq}) and then checks that the remaining equations are solved.

\bigskip

The general solution of the linearized equations of motion
can hence be expressed in terms of solutions to coupled second order equations or, equivalently, the fourth order equations as
\bea
a_{v} &=&  \frac{b}{2} \rho^{-z/2} (h_{(0) vv} + \rho h_{(2) vv}) + a_v^V; \\
a_{\rho} &=& \frac{1}{2} b \pa_{u}^{-1} h_{(2) uv} \rho^{-z/2} + a_{\rho}^V. \nn
\eea
Since $(a_v^V,a_{\rho}^V)$ satisfy coupled second order equations, the general solution involves four independent integration constants.
The other fluctuations can be formally expressed in terms of $(a_v^V,a_{\rho}^V)$ as
\bea
\pa_v a^{V}_{u} &=& - 4 \rho \pa_{\rho} a_{\rho}^V - \pa_u a_v^V + \sigma^2 \rho^{1-z} \pa_v a_{v}^{V}; \\
\pa_v a^{V}_u &\equiv& \pa_v a_u - b z \rho^{-z/2} \pa_{u}^{-1}  h_{(2) uv} - \frac{1}{2} b \rho^{-z/2} (\pa_u h_{(0)vv} + \rho \pa_v h_{(2) uv} ) \nn \\
\pa_{\rho}^2 h^{V}_{uu} &=& \pa_{\rho}^2 \left (  \rho^{1-z} \sigma^2 \pa_v^{-2} (z h_{(2)vv} + \pa_u \pa_v h_{(0)vv}) \right . \\
&& \qquad + \rho^{2-z} \frac{b^2 z}{2 (2-z)} h_{(2) uv} \nn \\
&& \qquad \left . + 2 z b \rho^{1-z} \pa_v^{-2} (\pa_{\rho}(\rho^{z/2} a_{v}^V) - \rho^{z/2} a_{\rho,v}^{V}) \right ). \nn
\eea
Now let us consider the differential equations satisfied by the vector fluctuations in more detail. If one Fourier transforms
to momentum space so that for every field $\phi(r,u,v)$
\be
\tilde{\phi}(r,k_u,k_v) = \int du dv e^{i k_u u + i k_v v } \phi(r,u,v),
\ee
then the operator $\Delta$ acts on $\tilde{\phi}$ as
\be
\Delta \tilde{\phi} = (\rho \pa_{\rho}^2 + \pa_{\rho} - \frac{z^2}{4 \rho} - \frac{1}{2} k_u k_v + \frac{\sigma^2}{4} \rho^{1-z} k_v^2)
\tilde{\phi}.
\ee
It is then natural to introduce a new dimensionless coordinate $x$
\be
x = (2 k_u k_v) \rho \equiv k^2 \rho,
\ee
such that
\be
\Delta \tilde{\phi} = k^2 ( x \pa_{x}^2 + \pa_{x} - \frac{z^2}{4 x} - \frac{1}{4} + \sigma^2 x^{1-z} k_{\chi} ) \tilde{\phi} \equiv k^2
\Delta_x \tilde{\phi},
\ee
where
\be
k_{\chi} \equiv 2^{z - 4} k_v^z k_u^{z-2}.
\ee
Then the fourth order equation for $a_{v}^{V}$ is
\be
\Delta_x^2 a_v^V = (z-1) \sigma^2 k_{\chi} x^{-z} a_v^V.
\ee
Since this equation only depends on the dimensionless coordinate $x$ and the quantity $\sigma^2 k_{\chi}$, the exact solution for $a_{v}^{V}$ can
only depend on these quantities, as discussed earlier.
Regularity throughout the spacetime will, as we show below, impose two conditions on the four independent solutions
of this equation. In what follows we will solve the equations at weak chirality, namely perturbatively in $\sigma^2$, in which case it is more
convenient to use the coupled second order equations rather than the fourth order equation.

\subsection{`T' and `X' modes of solution}

One can summarize the general solution of the linearized equations of motion as follows. The metric fluctuations are
\bea
h_{vv} &=& h_{(0) vv} + \rho h_{(2)vv}; \nn \\
h_{uv} &=& h_{(0)uv} + \rho h_{(2) uv} + \frac{b^2 z}{4} \left ( \frac{1}{(1-z)} \rho^{1-z} h_{(0) vv} + \frac{1}{(2-z)}
\rho^{2-z} h_{(2) vv} \right ); \label{summ} \\
h_{uu} &=& \td{h}_{(0) uu} + \rho \tilde{h}_{(2) uu} + \rho^{1-z} \sigma^2 \pa_v^{-2} (z h_{(2)vv} + \pa_u \pa_v h_{(0)vv}) \nn \\
&& \qquad + \rho^{2-z} \frac{b^2 z}{2 (2-z)} h_{(2) uv} \nn \\
&& \qquad + 2 z b \rho^{1-z} \pa_v^{-2} (\pa_{\rho}(\rho^{z/2} a_{v}^V) - \rho^{z/2} a_{\rho,v}^{V}). \nn
\eea
The vector fluctuations are
\bea
a_{v} &=&  \frac{b}{2} \rho^{-z/2} (h_{(0) vv} + \rho h_{(2) vv}) + a_v^V; \\
a_{\rho} &=& \frac{1}{2} b \pa_{u}^{-1} h_{(2) uv} \rho^{-z/2} + a_{\rho}^V; \nn \\
\pa_v a^{V}_u &=& \pa_v a_u - b z \rho^{-z/2} \pa_{u}^{-1}  h_{(2) uv} - \frac{1}{2} b \rho^{-z/2} (\pa_u h_{(0)vv} + \rho \pa_v h_{(2) uv }); \nn \\
\pa_v a^{V}_{u} &=& - 4 \rho \pa_{\rho} a_{\rho}^V - \pa_u a_v^V + \sigma^2 \rho^{1-z} \pa_v a_{v}^{V}. \nn
\eea
The propagating modes $a^{V}_{m}$ solve the coupled differential equations:
\be
\Delta a_v^V = \pa_v a^{V}_{\rho}; \qquad  \Delta a^{V}_{\rho} = \frac{\sigma^2}{4} (1-z) \rho^{-z} \pa_v a_v^V, \label{diff-eq}
\ee
where the second order differential operator $\Delta$ is given in \eqref{Delta}.

Let us express the source and normalizable modes in the asymptotic expansion of $h_{uu}$ as $\rho \rightarrow 0$ as
\be
h_{uu} = h_{(0) uu} + \rho h_{(2) uu} + \cdots,
\ee
respectively. These are given in terms of the integration constants $(\td{h}_{(0)uu}, \td{h}_{(2) uu})$ as
\bea
h_{(0) uu} &=& \td{h}_{(0)uu} + 2 z^2 b \pa_v^{-2} a^V_{(z)v}; \\
h_{(2) uu} &=& \td{h}_{(2) uu} + 2 z b (z+1) \pa_v^{-2} a^V_{(z+2) v} - 2 z b \pa_v^{-1} a^V_{(z) \rho}, \nn
\eea
where $a_{(m) a}^V$ is the coefficient of the term at order $\rho^{m/2}$ in the asymptotic expansion of $a^{V}_a$ as $\rho \rightarrow 0$.
Note that it is $h_{(0) uu}$ which is the source for dual operator, and
the $(u)$, $(v)$ and $(u v)$ components of Einstein equations at order $\rho^0$ enforce
the linearized Ward identities
\bea
\pa_v h_{(2) uv} &=& \pa_u h_{(2) vv}; \label{Ward-lin} \\
\pa_v {h}_{(2) uu} &=& \pa_u h_{(2) uv} - z b \pa_u a_{(z) v}^V, \nn \\
h_{(2) uv} &=& - \frac{1}{4} R[h_{(0)}] + \frac{z^2}{2} b a_{(z)v}^V, \nn
\eea
and
\be
R[h_{(0)}] = \pa_{u}^2 h_{(0)vv} + \pa_v^2 h_{(0) uu} - 2 \pa_{u} \pa_{v} h_{(0) uv}.
\ee
We have used the fact that
the coupled differential equations for $a^{V}_{m}$ can be solved asymptotically as $\rho \rightarrow 0$. The resulting solutions have the
structure expected from the previous (non-linear) analysis, namely
\bea
a^{V}_a &=& a^{V}_{(-z)a} \rho^{-z/2} + \cdots + a^{V}_{(z) a} \rho^{z/2} + \cdots, \\
a^{V}_{\rho} &=& \frac{1}{2z} (\pa_v a^{V}_{(-z)u} + \pa_{u} a^{V}_{(-z)v}) \rho^{-z/2} + \cdots \nn \\
&& \qquad - \frac{1}{2z} (\pa_v a^{V}_{(z)u} + \pa_{u} a^{V}_{(z)v}) \rho^{z/2} + \cdots \nn
\eea
where we have isolated the terms corresponding to the operator source ($a^{V}_{(-z)a}$) and operator expectation value ($a^{V}_{(z)a}$)
respectively. The analogous terms in the radial component of the vector field are completely determined in terms of these components.
The case of $z=2$ is special, as the radial powers in the independent solutions depend explicitly on $b^2$, see \cite{Guica:2010sw}, because
the dimension of the dual operator is modified at non-zero $b$.

Recall that for $z< 1$ the holographic one point functions at the linearized level are given in terms of coefficients in the asymptotic expansion as
\bea
\< T_{vv} \> &=& \frac{1}{\kappa^2} h_{(2) vv}; \qquad
\< T_{uv} \> = - \frac{1}{\kappa^2} h_{(2) uv}; \label{vev-lin} \\
\< T_{uu} \> &=& \frac{1}{\kappa^2} \left ( h_{(2)uu} - b z a_{(z)u} \right ); \nn \\
\< V_a \> &=& - \frac{z}{\kappa^2} a_{(z) a}. \nn
\eea
Combining the propagating solution for $a^{V}$ with the other modes results in the following source and
vev terms in the asymptotic expansions for the metric and vector fluctuations
\bea
h_{vv} &=& h_{(0)vv} + \rho h_{(2) vv}; \nn \\
h_{uv} &=& h_{(0)uv} + \rho h_{(2) uv} + \cdots \nn \\
h_{uu} &=& h_{(0)uu} + \rho {h}_{(2) uu} + \cdots \\
a_v &=& \rho^{-z/2}
\left( a^{V}_{(-z)v} + \frac{b}{2} h_{(0) vv} \right ) + \cdots + a^{V}_{(z)v} \rho^{z/2} + \cdots \nn \\
a_{u} &=&  \rho^{-z/2} \left ( a^{V}_{(-z)u} + \frac{b}{2} \pa_{v}^{-1} \pa_{u} h_{(0) vv} + b z \pa_{v}^{-1}
\pa_{u}^{-1} h_{(2) uv} \right ) \nn \\
&& \qquad + a^{V}_{(z)u} \rho^{z/2} + \cdots \nn
\eea
These expressions imply that the stress energy tensor sources are $g_{(0)ab} = \eta_{(0)ab} + h_{(0)ab}$ and
the vector operator sources are given by
\bea
b_{(-z)v} &=& a_{(-z) v} = \left( a^{V}_{(-z)v} + \frac{b}{2} h_{(0) vv} \right ); \label{vec-source} \\
b_{(-z)u} &=& b + a_{(-z) u} = b + \left ( a^{V}_{(-z)u} + \frac{b}{2} \pa_{v}^{-1} \pa_{u} h_{(0) vv} + b z \pa_{v}^{-1}
\pa_{u}^{-1} h_{(2) uv} \right ). \nn
\eea
Thus in particular the fluctuation $h_{(0)ab}$ sources not just the stress energy tensor but also the vector operator.
To compute two point functions of the
stress energy tensor one should set the vector source to zero, by switching on appropriate $a^{V}_{(-z) a}$, whilst to compute the two point functions
of the vector operator one should set to zero $h_{(0) ab}$. Note that switching off the sources for either set of operators does not switch off their
expectation values, since the two point functions in the deformed theory are non-diagonal.

In \cite{Guica:2010sw} the general linearized solution for $z=2$ was given in terms of independent solutions
of the equations of motion, the `T' and `X' modes. The `T' mode solution is the $z \rightarrow 2$ limit of the
solution given above with $a^{V} =0$, which involves only the integration constants $(h_{(0) ab}, h_{(2) ab})$. The limit
of $z \rightarrow 2$ requires
\be
\frac{1}{(2-z)} \rho^{2-z} \rightarrow \ln (\rho).
\ee
This `T' mode solution is non-dynamical, in that there is no bulk differential equation satisfied by these modes. From a field theoretic
perspective, these correspond to quantities which are completely determined by Ward identities. From the bulk perspective the corresponding statement
is that the `T' mode solution is equivalent to a bulk diffeomorphism. To show this, let us consider a bulk diffeomorphism generated by a vector field
$\zeta_{m}$ such that
\be
\delta g_{mn} = (D_m \zeta_n + D_n \zeta_m).
\ee
Restricting to diffeomorphisms which respect the Fefferman-Graham form of the metric requires $\delta g_{rr} = \delta_{r a} = 0$ and hence
\bea
\zeta_{\rho} &=& \frac{\zeta}{\rho}; \qquad
\zeta_{v} = \frac{\zeta_{(0) v}}{\rho} - \pa_v \zeta; \\
\zeta_{u} &=&  \frac{\zeta_{(0) u}}{\rho} - \pa_v \zeta + \sigma^2 \rho^{-z} \zeta_{(0) v} \nn \\
&& \qquad + \frac{\sigma^2 (z-1)}{(2-z)} \rho^{1 - z} \pa_v \zeta, \nn
\eea
where $(\zeta,\zeta_{(0) a})$ are independent arbitrary functions of $(u,v)$. The metric variations are then
\bea
\delta g_{vv} &=& \frac{1}{\rho} \left ( 2 \pa_v \zeta_{(0)v} - 2 \pa_v^2 \zeta \rho \right ); \\
\delta g_{uu} &=& \frac{1}{\rho} \left ( 2 \pa_u \zeta_{(0)u} - 2 \pa_u^2 \zeta \rho \right ) \nn \\
&& \qquad - \frac{4 \sigma^2}{(z-1)} \rho^{-z} \zeta + \frac{2 \sigma^2}{(2-z)} \rho^{1-z} \pa_u \pa_v \zeta; \nn \\
\delta g_{uv} &=&  \frac{1}{\rho} \left ( (\pa_u \zeta_{(0) v} + \pa_v \zeta_{(0) u}) - 4 \zeta - 2 \pa_u \pa_v \zeta \rho \right ) \nn \\
&& \qquad + \sigma^2 \rho^{-z} \pa_v \zeta_{(0) v} + \frac{\sigma^2 (z-1) }{(2-z)} \rho^{1 - z} \pa^2_v \zeta, \nn
\eea
and using the analog of \eqref{diffeo} the vector field fluctuations are
\bea
\delta b_{\rho} &=& - b \pa_v \zeta \rho^{-z/2}; \\
\delta b_{v} &=& b \rho^{-z/2} (\pa_v \zeta_{(0)v} - \rho \pa_v^2 \zeta); \nn \\
\delta b_{u} &=& - b \rho^{-z/2}  (\pa_{u} \zeta_{(0)v} - \rho \pa_v \pa_u \zeta) - 2 b z \rho^{-z/2} \zeta. \nn
\eea
Noting that $\delta g_{mn} = h_{mn}/\rho$ this agrees with the `T' mode fluctuations, under the identifications
\be
h_{(0) vv} = 2 \pa_v \zeta_{(0)v}; \qquad
h_{(0) uu} = 2 \pa_u \zeta_{(0)u}; \qquad
h_{(0) uv} = (\pa_u \zeta_{(0) v} + \pa_v \zeta_{(0) u}) - 4 \zeta,
\ee
with all other modes determined in terms of these quantities. The `X' mode solution of \cite{Guica:2010sw} corresponds to our
propagating solution $a^{V}$. In the limit of $z \rightarrow 2$ the
coupled differential equations (\ref{diff-eq}) remain well-defined, but the asymptotic solutions of these equations depend explicitly
on $b^2$, since the corresponding vector operator picks up an anomalous dimension at non-zero $b$ \cite{Guica:2010sw}.

One of the puzzling features in \cite{Guica:2010sw} was that in
the `T' mode solution the vector field is expressed non-locally in terms
of the ``source" data $h_{(0) ab}$. In the case of $z < 1$, where the relationship between asymptotics of the fluctuations and operator data is known,
the reason for this feature is now clear: $h_{(0) ab}$ does not source just $T_{ab}$, but it also sources the vector operator $V_{a}$. Moreover, from
\eqref{vec-source}, one sees that the source $a_{(-z) u}$ is non-locally expressed in terms of $h_{(0)ab}$.
Note however that the `T' mode solution is manifestly local when expressed in terms
of the vector $\zeta^m$ generating the bulk diffeomorphism. From the boundary perspective $\zeta$ parameterizes a Weyl rescaling, whilst $\zeta_{(0)a}$
generates a boundary diffeomorphism.

Since the asymptotic expansion is local in the $\zeta_{m}$ it would be natural to set up a variational problem in terms of these quantities. Such a vector field
formalism for the case of $z > 1$ will be explored elsewhere. In the case of $z < 1$, it is not necessary to use such a formalism as
one can exploit the fact that the spacetime is asymptotically
locally anti-de Sitter to set up the variational problem and holographic renormalization in terms of the usual data $(g_{(0) ab}, b_{(-z) a})$.
In this case one can compute the two point functions as follows: using the Ward identities the only undetermined information is the two
point functions of the vector operator. These can be computed by setting to zero the sources $g_{(0) ab}$, and solving the differential equations
for the propagating modes $a^{V}$. The sources for the vector operator will induce expectation values for the stress energy tensor, corresponding to
the cross correlators between the stress energy tensor and the vector operators. These two point functions are also completely
determined by the Ward identities, and therefore
do not give additional information. By the arguments given in the previous section, the same procedure may be carried out
for generic $z > 1$, when $z$ is not rational, and when the deforming vector operator does not acquire an anomalous dimension.
What remains to be done, therefore, is to find the regular solutions for $a^{V}$.

\subsection{Solution around $AdS_3$}

Let us first solve the vector field equations for $\sigma^2 =0$. Using (\ref{2nd1}) and (\ref{2nd1}), one can show that the regular solutions are:
\bea
a_{v} (\rho,k) &=& a_{(-z)v} (k) \td{K}_{z} (k \sqrt{\rho}) +  \frac{2 i k_v}{k^{1- z}} a_{(-z) \rho} \frac{2^{1- z} \sqrt{\rho}}{\Gamma(z)}
{K}_{(z-1)} (k \sqrt{\rho}); \nn \\
a_{\rho} (\rho,k) &=&  a_{(-z) \rho} \td{K}_{z} (k \sqrt{\rho}), \label{n=0}
\eea
where $\td{K}_{z} (k \sqrt{\rho})$ represents the modified Bessel function with a specific normalization such that
\bea
\td{K}_{z} (k \sqrt{\rho}) &=& \frac{2^{(1- z)}}{\Gamma(z)} k^{z} K_{z}(k \sqrt{\rho}); \\
&=& \rho^{-z/2} (1 + \frac{k^2 \rho}{4(1-z)} + \cdots) + \frac{\Gamma(- z)}{2^{2 z} \Gamma(z)} k^{2 z} \rho^{z/2}
(1 + \frac{k^2 \rho}{4(1+ z)} + \cdots),
\eea
and the latter is the expansion as $\rho \rightarrow 0$. In solving the equations recurrence relations for modified Bessel functions are useful:
\be
K_{z + 1} (x) = K_{z -1} (x) + \frac{2 z}{x} K_{z} (x); \qquad
\pa_x K_{z}(x) = - K_{z -1} (x) - \frac{z}{x} K_{z} (x).
\ee
Using \eqref{lin-div-eq} one can then show that the solution
\bea
a_{u} (\rho,k) &=&  a_{(-z)u} (k) \td{K}_{z} (k \sqrt{\rho}) + \frac{2 i k_u}{k^{1-z}} a_{(-z) \rho} \frac{2^{1- z} \sqrt{\rho}}{\Gamma(z)}
{K}_{(z-1)} (k \sqrt{\rho}) \nn \\
a_{(-z) \rho} &=& \frac{1}{2z} (\pa_v a_{(-z) u} + \pa_u a_{(-z) v}),
\eea
satisfies all remaining equations. As a consistency check note that the asymptotic expansions of all
vector field components as
$\rho \rightarrow 0$ agree with those given in section \ref{sec:hr}, with the normalizable modes
determined in terms of the non-normalizable modes as follows
\bea
a_{v} (\rho,k) &=& a_{(-z)v} (k) \rho^{-z/2} + \cdots \frac{k_v k^{2z}}{k_u}
\frac{\Gamma(-z)}{2^{2z} \Gamma(z)} a_{(-z) u} \rho^{z/2} + \cdots \\
a_{u} (\rho,k) &=& a_{(-z)u} (k) \rho^{-z/2} + \cdots \frac{k_u k^{2z}}{k_v}
\frac{\Gamma(-z)}{2^{2z} \Gamma(z)} a_{(-z) v} \rho^{z/2} + \cdots \nn
\eea
The two point functions are computed using
\be
\< V_{a} (k) V_{b} (-k) \> =
\frac{z}{\kappa^2} \frac{\delta b_{(z)a}}{\delta b_{(-z)}^b} = \frac{z}{\kappa^2} \frac{\delta a_{(z)a}}{\delta a_{(-z)}^b} + \cdots\, ,
\ee
where the ellipses denotes contact terms, and thus
\be
\< V_{v} (k) V_{v} (-k) \> = - \frac{1}{\kappa^2} \frac{k_v k^{2z}}{k_u} \frac{\Gamma(1-z)}{2^{2z} \Gamma(z)}; \qquad
\< V_{u} (k) V_{u} (-k) \> = - \frac{1}{\kappa^2} \frac{k_u k^{2z}}{k_v} \frac{\Gamma(1-z)}{2^{2z} \Gamma(z)},
\ee
with the cross correlation function vanishing, as it should, since the operators have different scaling weights.
These expressions can be written in position space as follows. Recall that
the general expression for the Fourier transform of a polynomial in $d$ dimensions is
\be
\frac{1}{(2 \pi)^d} \int d^d k e^{-i \vec{k} \cdot \vec{x}} (k^2)^{\lambda} =
\pi^{-d/2} 2^{2 \lambda} \frac{\Gamma(d/2 + \lambda)}{\Gamma (-\lambda)} (|x|^2)^{-\lambda - d/2}, \label{ft-imp}
\ee
which is valid when $\lambda \neq - (d/2 + n)$, where $n$ is zero or a positive integer. Using this Fourier transform,
and its derivatives with respect to $x$, one obtains
\be
\< V_{v} (x) V_{v} (0) \> = \frac{z (z+1)}{4 \pi \kappa^2} \frac{1}{|x|^{2z} v^2}; \qquad
\< V_{u} (x) V_{u} (0) \> = \frac{z (z+1)}{4 \pi \kappa^2} \frac{1}{|x|^{2z} u^2},
\ee
which is of the expected form for operators of these scaling dimensions.

\subsection{General solution}

We now consider the case of $\sigma \neq 0$ with the sources for the dual stress-energy tensor switched off.
Let us first express the asymptotic expansions of the solutions to the dynamical vector field equation as
\be
a^V_{a} = \rho^{-z/2} a^V_{(-z)a} + \cdots
+ X_{a}^{\;b} (\sigma,k) a^{V}_{(-z) b} \rho^{z/2} + \cdots,
\ee
where the matrix $X_{a}^{\;b} (\sigma,k)$ is to be determined by solving the inhomogeneous differential
equations exactly and imposing regularity conditions. The asymptotic expansion of the vector field is
then written in terms of this data as
\be
a_{a} = \rho^{-z/2} (a^V_{(-z)a} + b\, z\, \delta_{au} \pa_v^{-1} \pa_u^{-1} h_{(2)uv} ) + \cdots
+ X_{a}^{\;b} (\sigma,k) a^{V}_{(-z) b} \rho^{z/2} + \cdots,
\ee
Note that the source for the vector operator includes another term involving $h_{(2)uv}$, as
the latter is not automatically set to zero by setting $h_{(0) ab} = 0$. Indeed
from the linearized Ward identity \eqref{Ward-lin} one knows that
\be
h_{(2) uv} = \frac{z^2}{2} b a^{V}_{(z) v} \equiv \frac{z^2}{2} b X_{v}^{\;b} (\sigma,k) a^{V}_{(-z)b}.
\ee
The true vector operator sources are thus defined in terms of the asymptotic solutions to the dynamical
equations as
\be
a_{(-z)a} = a_{(-z)a}^V  -  \frac{b^2 z^3}{k^2} \delta_{au} X_{v}^{\;b} (\sigma,k) a^{V}_{(-z)b},
\ee
and therefore
\bea
a_{(-z) v} &=& a_{(-z) v}^V; \\
a_{(-z) u} &=& (1 - \frac{b^2 z^3}{k^2} X_{vv} ) a_{(-z) u}^V - \frac{b^2 z^3}{k^2} X_{u}^{\;v} a^V_{(-z) v}. \nn
\eea
These relations allow one to rewrite the modes $a_{(-z)a}^{V}$ in terms of the true sources, and thence one
can also obtain the relationship between the normalizable modes $a_{(z)a}^{V}$ and the sources.
Functionally differentiating the linearized one point functions with respect to the sources one then finds that
\bea
\< V_{v} V_{v} \> &=& \frac{z}{\kappa^2} \left ( 1 - \frac{b^2 z^3}{k^2} X_{vv} \right )^{-1} X_{vv}; \\
\< V_{u} V_{v} \> &=& \frac{z}{\kappa^2} X_{vu}
\left (1 - \frac{b^2 z^3}{k^2} X_{vv} \right )^{-1} ; \nn \\
& \equiv & \frac{z}{\kappa^2} X_{u v} \left (1 - \frac{b^2 z^3}{k^2} X_{vv} \right )^{-1}; \nn \\
\< V_{u} V_{u} \> &=& \frac{z}{\kappa^2} \left ( X_{uu}  + \frac{b^2 z^3}{k^2} X_{uv} X_{vu}
(1 - \frac{b^2 z^3}{k^2} X_{vv})^{-1} \right ). \nn
\eea
The fact that $\< V_{v} V_{u} \> = \< V_{u} V_{v} \>$ must then follow from the symmetry of the matrix $X_{ab}$ that
arises in solving the differential equations.

Next let us consider the vector field equations at $\sigma \neq 0$. It is useful to write the equations \eqref{2nd1}
and \eqref{2nd2} in the form:
\bea
\Delta_0 a_v^V - \pa_v a_{\rho}^V &=& \frac{1}{4} \sigma^2 \rho^{1-z} \pa_v^2 a^{V}_v; \\
\Delta_0 a_{\rho}^V &=& \frac{1}{4} \sigma^2 (1-z) \rho^{-z} (\rho \pa_v^2 a_{\rho}^V + \pa_v a_v^V), \nn
\eea
where $\Delta_0$ is the restriction of the differential operator $\Delta$ to $\sigma^2 = 0$. It is interesting
to note that the corrections to the differential equation at $\sigma^2 \neq 0 $ vanish when the lightcone momentum $k_{v} = 0$.
Working in conformal perturbation theory we noted that corrections were organized in powers of $\sigma^2 k_v^2$, and the same behavior is found
holographically. Let us try to solve
the equations perturbatively in $\sigma k_v$ at $k_v \neq 0$ by looking for solutions of the form
\bea
a_{v}^{V} &=& \sum_{n > 0} \sigma^{2 n} (a_{v}^{V})_{n}; \\
a_{\rho}^{V} &=& \sum_{n > 0} \sigma^{2 n} (a_{\rho}^{V})_{n}, \nn
\eea
where the $n=0$ solutions are given by \eqref{n=0}. The coupled differential equations then reduce to pairs of inhomogeneous
differential equations generating a recurrence relation
\bea
\Delta_0 (a_v^V)_{n+1} - \pa_v (a_{\rho}^V)_{n+1} &=& \frac{1}{4} \rho^{1-z} \pa_v^2 (a^{V}_v)_n; \\
\Delta_0 (a_{\rho}^V)_{n+1} &=& \frac{1}{4} (1-z) \rho^{-z} (\rho \pa_v^2 (a_{\rho}^V)_n + \pa_v (a_v^V)_n). \nn
\eea
For generic values of $z$ the corrections $(a_{a}^V)_n$ are bounded as $\rho \rightarrow \infty$, since the differential operator $\Delta_0$ has an essential
singularity as $\rho \rightarrow \infty$ and so the regular $n=0$ solutions decay exponentially there:
\be
a_{\rho} (\rho,k) =  a_{(-z) \rho} e^{- k \sqrt{\rho}} \left (
\frac{(k/2)^{z - 1/2} \sqrt{\pi}}{\Gamma(z) \rho^{1/4}} + {\cal O} (\rho^{-3/4}) \right ).
\ee
Solving for the inhomogeneous contributions to the corrections as $\rho \rightarrow \infty$ one finds that they also behave as
\be
(a_{a}^V)_n \sim e^{- k \sqrt{\rho}} \rho^{-1/4},
\ee
and are hence exponentially small.

Once we have established that the inhomogeneous contributions to the corrections are finite everywhere,
we need to solve the inhomogeneous differential equations to extract the asymptotic coefficients $(a^{V}_{(-z)a},a^{V}_{(z) a})$. This could
be carried out numerically for finite chirality $b^2$, and can be done
perturbatively in $b^2$ at small chirality, using the Green function for the differential operator $\Delta_0$, which is given in the appendix.
This results in the following correlation functions
\bea
\< V_{v} (k) V_{v} (-k) \> &=& - \frac{1}{\kappa^2} \frac{k_v k^{2z}}{k_u} \frac{\Gamma(1-z)}{2^{2z} \Gamma(z)} \left (1 + c_{vv} k_{\chi}^2 \sigma^2 \right);
\nn \\
\< V_{v} (k) V_{u} (-k) \> &=& - \frac{1}{\kappa^2} c_{vu} k_{\chi}^2 \sigma^2; \label{point} \\
\< V_{u} (k) V_{u} (-k) \> &=& - \frac{1}{\kappa^2} \frac{k_u k^{2z}}{k_v} \frac{\Gamma(1-z)}{2^{2z} \Gamma(z)} \left (1 + c_{uu} k_{\chi}^2 \sigma^2 \right ), \nn
\eea
where the constant numerical coefficients $c_{ab}$ are given in the appendix.

To summarize,
these correlation functions are sufficient to reconstruct all two point functions of the stress energy tensor and vector operator, to leading order in $b^2$. The
functional form of the correlation functions is as anticipated from anisotropic scale invariance and conformal perturbation theory. As we will emphasize in the
conclusions, the holographic models for scale invariance with exponent $z$ always include fields dual to the deforming, Lorentz symmetry breaking, operators.
These operators must therefore necessarily play an important r\^{o}le in the physics of the condensed matter system being modeled. At small chirality, the correlation
functions of these operators are given by the above formulae and these should match the features of the system under consideration. One would also like
the finite temperature holographic realization to match the behavior of the physical system under consideration, and we will next turn to modeling finite temperature
physics with black holes.

\section{Black holes} \label{se:blackhole}

It would be interesting to find black hole solutions of the gravity-vector system, in order to probe the phase structure of the anisotropic theory. Again
there will be a qualitative difference between the cases of $z < 1$ and $z > 1$. In the former case, the deformation is relevant with respect to
the conformal symmetry and one would only expect to retain the effects of the deformation at temperatures which are small compared to the deformation
parameter:
\be
T \ll b^{1/z}.
\ee
Let us start by considering the following black hole solution in three dimensions
\bea
ds^{2} &=&
\frac{dr^2}{r^2 (1 - ( r/r_{+})^{2(2-z)})} \label{black-hole} + \\
&& \qquad \frac{1}{r^2} \left ( - \left( 2 - (r/r_{+})^{2(1-z)} - (r/r_{+})^{2} \right) d \eta^{2}
+ 2 \left( 1- (r/r_{+})^{2} \right) d\eta dx + (r/r_{+})^{2} dx^{2} \right). \nn
\eea
This geometry describes a black hole with Killing horizon $r = r_{+}$ and generator
$K = \partial_{\eta}$. The critical exponent is restricted to $z < 2$, otherwise $g_{rr} \not \to 1/r^{2}$ as $r \to 0$.
Under the coordinate transformation: $x := \x + \eta$, the metric becomes:
\be
ds^{2} = {1 \over r^{2}} \left( {dr^{2} \over 1 - (r/r_{+})^{2(2-z)}} + (r/r_{+})^{2(1-z)} d\eta^{2} + 2\, d\x\, d\eta + (r/r_{+})^{2} d\x^{2} \right).
\ee
In these coordinates, the black hole is manifestly asymptotic to the chiral scale-invariant background as $r \to 0$ for $z < 2$. Next one lets
$\eta = \sigma r_{+}^{1-z} u$ and $\x = r_{+}^{z-1} v/\sigma$ so that
\be
ds^2 = \frac{dr^2}{r^2 (1 -  (r/r_{+})^{2(2-z)})} + \sigma^2 r^{-2z} d u^2 + \frac{2}{r^2} d u dv + \frac{d v^2}{\sigma^2 r_{+}^{4 -2 z}}.
\ee
The anisotropic scale invariant background can be obtained as the
zero temperature limit of the black hole, corresponding to $r_{+} \rightarrow \infty$ with $\sigma$ finite.

Einstein equations admitting such solutions can be constructed as follows. Writing the scale invariant geometry as
\be
ds^{2} = {1 \over r^{2}} \left( dr^{2} + \s^{2} r^{2(1-z)} du^{2} + 2\, d\x\, du \right)
\ee
note that the Einstein tensor $G_{ab}$ satisfies
\be
G_{ab} = g_{ab} + z^{2} B_{a} B_{b}, \qquad  B = b r^{-z} du \qquad b^{2} = 2 { 1-z \over z } \s^2.
\ee
The contravariant components are correspondingly
\be
G^{ab} = g^{ab} + z^{2} B^{a}B^{b}, \qquad B = -b r^{2-z} \partial_{\x}.
\ee
For the above black hole solution, the Einstein tensor satisfies
\be
G^{ab} = g^{ab} + z^{2} B^{a}B^{b}, \qquad B = -b r^{2-z} \partial_{\x}, \qquad
b^{2} = 2 { 1-z \over z } r_{+}^{-2(1-z)}.
\ee
This means that the contravariant energy tensor is actually exactly the same in both cases.
However, while for the scale invariant background one can write a pure
Proca action generating the required field equations, for the black hole solution one cannot.
Note that the case of $z = 1$ is exceptional: the above black hole reduces to the BTZ black hole which satisfies
the Einstein equations without matter. The case of $z=0$ in three dimensions is also special, as the spacetime is Einstein.

The fact that the black hole solution does not follow from a Proca action suggests that a string theory embedding
may give rise to consistent truncations involving not just vectors, but vectors coupled to scalars. This is indeed known
to be the case for Schr\"{o}dinger ($z=2$) in five bulk dimensions, see the consistent truncation found in \cite{Maldacena:2008wh}.
It is interesting however to note that the black hole solution can be supported
by dust or by a perfect fluid; appropriate actions for dust solutions can be found in \cite{Brown:1994py} and for perfect fluids in \cite{Brown:1994ix}.

Starting from \eqref{black-hole}, the normal to the hypersurfaces of constant $r$
is null at the horizon as well as the Killing vector $\partial_{\eta}$: $\ ||dr||^{2} = 0 = ||\partial_{\eta}||$ at $r_{+}$.
To see that the Killing is normal to $r = r_{+}$, one rewrites \eqref{black-hole} in the form:
\be
ds^{2} = {1 \over r^{2}} \left[ - \left( 2 - (r/r_{+})^{2(1-z)} - (r/r_{+})^{2} \right) \left( d\eta^{2} - dr^{*2} \right)
+ 2 \left( 1 - (r/r_{+})^{2} \right) d\eta dx + (r/r_{+})^{2} dx^{2} \right],
\ee
where
\be
dr^{*} = {dr \over \sqrt{1 - (r/r_{+})^{2(2-z)}} \sqrt{2 - (r/r_{+})^{2(1-z)} - (r/r_{+})^{2}}}.
\ee
Then define: $\eta = U + r^{*}$
\bea
ds^{2} &=&
{1 \over r^{2}} \left[ - \left( 2 - (r/r_{+})^{2(1-z)} - (r/r_{+})^{2} \right) dU^{2}
- 2 \sqrt{ {2 - (r/r_{+})^{2(1-z)} - (r/r_{+})^{2} \over 1 - (r/r_{+})^{2(2-z)}} } dU dr \right. \nn \\
&& + 2 \left( 1 - (r/r_{+})^{2} \right) ( dUdx + dr^{*}dx ) + (r/r_{+})^{2} dx^{2} \Bigg]. \label{hor}
\eea
In this coordinate system, the metric is well behaved at the horizon with
the metric close to the horizon being
\be
ds^{2} = {1 \over r^{2}} \left( - 2\, dUdr + {1 \over 2-z}\, dxdr + dx^{2} \right),
\ee
which is well behaved everywhere near the horizon. One can further define $U = \hf y + {x \over 2(2-z)}$ with $r = {1 \over R}$
to obtain
\be
ds^{2} = dydR + R^{2}dx^{2},\qquad (r \to r_{+}).
\ee
This is exactly the same metric as that of the non-rotating BTZ black hole near the
horizon in Eddington-Finkelstein coordinates as long as one compactifies the coordinate $x$ with period $2\, \pi$.
From \eqref{hor}, the Killing vector $k = \partial_{\eta}$ in this coordinate system becomes $\partial_{U}$. This means that
\be
\bar{k} = g_{ac} k^{c} dx^{a} = g_{UU} dU + g_{rU} dr + g_{xU} dx.
\ee
At the horizon: $\bar{k} = {1 \over r_{+}^{2}}\, dr$, which implies that the horizon is indeed
a Killing horizon with respect to $\partial_{\eta}$.

The temperature of the black hole \eqref{black-hole} is
\be
T_L = {\kappa \over 2 \pi} = {2-z \over 2 \pi r_{+}},
\ee
where $\kappa$ is the surface gravity, whilst the entropy is given by
\be
S = \frac{\beta_x}{4 G_3 r_{+}},
\ee
where $\beta_{x}$ is the periodicity of the $x$ direction and $G_3$ is the Newton constant. The entropy
density $s = S/\beta_{x}$ can be expressed as
\be
s = \frac{\pi}{2 G_3 (2-z)} T_L,
\ee
the form of which is determined on dimensional and scaling grounds.

\bigskip

In the absence of a complete solution involving appropriate fields, one cannot directly interpret the black hole
\eqref{black-hole} in terms of finite temperature behavior of the deformed chiral theory. However, one can make
interesting preliminary observations: the temperature is associated with the periodicity of the Euclidean coordinate
$\bar{u} = i u$. Note however that $u$ is a null coordinate in the quantum field theory, and therefore
the temperature $T_L$ relates to that in the left moving sector of the field theory, hence the notation used. It would be interesting
to find an explicit embedding of this black hole into string theory, and thence its interpretation as a thermal state in
the dual field theory.

\section{Conclusions} \label{se:conclusions}

In this paper we have explored features of theories holographically dual to chiral scale invariant geometries \eqref{eq:nrmetric}.
The dual theories are anisotropic but scale invariant deformations of $d$-dimensional conformal field theories in a flat background with
coordinates $(u,v,x^i)$. Dimensional reduction along the $u$ coordinate for $z < 1$ and along the $v$ coordinate for $z > 1$ results
in a $(d-1)$-dimensional theory with non-relativistic scale invariance. This reduction introduces technical issues, since it is along a null
direction, but from the perspective of CMT applications it leads to an important conceptual question: {\it to what extent can the physics of the CMT
system actually be captured by the DLCQ of a theory in one higher dimension?}

In the holographic realizations of anisotropic systems discussed here, a central r\^{o}le is played by the exactly marginal deforming vector
operator ${\cal V}_v$. It
is this operator which breaks the relativistic conformal symmetry, and at any value of the deformation parameter $b$ the operator will remain exactly
marginal with respect to the anisotropic symmetry. In the $d=2$ system discussed by Cardy in \cite{Cardy:1992tq}, the deforming operator indeed had an
interpretation in the physical system being modelled: the Hamiltonian included a chiral interaction, which remained finite in the continuum
limit at criticality.

By contrast, suppose
one considers the reduction of the deformed theory along a null direction to obtain a non-relativistic theory in $(d-1)$ dimensions, focusing
on the case of $z > 1$, in which case $v$ is the appropriate null reduction. Then, the deformed theory in the lower dimension can be expressed as:
\be
S_{\rm cft} + \int du dv d^{d-2}x b {\cal V}_{v} + \cdots \rightarrow S_{\rm red} + \int du d^{d-2} x b {\cal V}_v(k_v = 0) + \cdots,
\ee
where the reduction of the
original CFT action, $S_{\rm red}$, can be formally decomposed into a sum over Kaluza-Klein harmonics of different discrete lightcone momenta $k_v$
and the deformation involves only the $k_v = 0$ harmonic. Note that the null reduction of the $d$-dimensional CFT will not have $(d-1)$-dimensional conformal
invariance. From the perspective of the lower-dimensional theory, ${\cal V}_{v} (k_v = 0)$ is a scalar
operator with ``charge" $k_v  = 0$. The ellipses denote terms higher order in $b$. If the holographic theory is to provide a good description of a
physical system, such as fermions at unitarity, then there must be a r\^{o}le for the deforming operator in that system. The parameter $b$ should characterize
a line of fixed points of the system and the physical features of the correlation functions of the deforming operator should also match.

Given the general framework for the holographic duality developed here, it would be interesting to explore whether any non-relativistic scale invariant systems do
admit this kind of exactly marginal deformations. It seems likely that many interesting non-relativistic scale-invariant CMT systems will not have these
features, and will need to be modeled in a different holographic manner (without the additional null direction). Even if this turns out to be the case, the
geometries \eqref{eq:nrmetric} without null compactifications may be useful in describing $d$-dimensional anisotropic systems such as those explored in
\cite{Cardy:1992tq}.

Finally, let us comment on the relation between this work and recent attempts to understand the microscopics of the Kerr/CFT correspondence,
\cite{Compere:2010uk} and \cite{Guica:2010ej}. In \cite{Compere:2010uk} a supergravity solution was considered which interpolates between a
self-dual null orbifold of $AdS_3 \times S^2$ and the near-horizon limit of the extremal Kerr black hole times a circle, and this
interpolation was used to understand the holographic dual, in terms of a deformation of the DLCQ of the MSW conformal field theory. In particular, the deforming
operators are vector operators, which respect anisotropic scale invariance; the leading deformation being a $(2,1)$ operator in the CFT, which is $z=2$
in our classification. It would be very interesting to explore further whether such anisotropic scale-invariant deformations of a CFT can
be used to provide a holographic description of extremal Kerr solutions.

\section*{Acknowledgments}

MMT would like to thank the Galileo Galilei Institute
for Theoretical Physics and the Simons Workshop in Mathematics and Physics 2010
for hospitality during the completion of this work. RCC acknowledges the support from the
Funda\c{c}\~{a}o para a Ci\^{e}ncia e Tecnologia (FCT, Portugal) via the grant SFRH/BD/43182/2008.
This work is part of the research program of the `Stichting voor
Fundamenteel Onderzoek der Materie (FOM)', which is financially
supported by the `Nederlandse Organisatie voor Wetenschappelijk
Onderzoek (NWO)'. The authors acknowledge support from NWO
via the Vidi grant ``Holography, duality and time
dependence in string theory".

\section{Appendix: Solution of vector equations}

The homogeneous equation
\be
\left (\rho \pa_{\rho}^2 + \pa_{\rho} - \frac{z^2}{4 \rho} - \frac{k^2}{4} \right ) \phi (\rho) = 0,
\ee
admits modified Bessel functions as solutions
\be
\phi = \alpha I_{z} (k \sqrt{\rho}) + \beta K_{z} (k \sqrt{\rho}).
\ee
The solution which is regular as $ \rho \rightarrow \infty$ is the second, so $\alpha = 0$.

Let us next consider a generic inhomogeneous equation
\be
\left (\rho \pa_{\rho}^2 + \pa_{\rho} - \frac{z^2}{4 \rho} - \frac{k^2}{4} \right ) \phi (\rho) =  g(\rho).
\ee
By defining $x = k\sqrt{\rho}$, it becomes:
\be
\Delta_{x} \phi(x) 
= \left[ \partial_{x} \left( x\, \partial_{x} \right) - \left( x + {z^{2} \over x} \right) \right] \phi(x) = x \left( {2 \over k} \right)^{2} g([x/k]^{2}) := h(x). 
\ee
The general solution to this equation is
\begin{equation}
\phi(x) = \phi_{0}(x) + \int_{0}^{\infty} dx' G(x,x') h(x'),
\end{equation}
where $\phi_{0}(x)$ satisfies the homogeneous equation (the regular solution throughout the bulk being $K_{z}(x)$)
and the Green's function is defined by
\be
\Delta_{x} G(x,x') = \delta(x'-x).
\ee
Then, the solution for the Green's function for $x\neq x'$ is:
\begin{equation}
G(x',x) =
\begin{cases}
A(x') K_{z}(x)\quad :\quad x > x'\\
B(x') I_{z}(x)\quad :\quad x < x'
\end{cases}
\end{equation}
For $x > x'$, one chooses the $K_{z}(x)$ so that the Green's function is regular as $x \to \infty$. For $x < x'$, one chooses the $I_{z}(x)$ so that the 
results for the case $\sigma^{2} = 0$ are recovered. Note that $I_{z}(x)$ does not contain the $x^{-z}$ power, only the $x^{z}$ one.
In order to find the coefficients, one imposes continuity in the Green's function and integrating the equation for $G(x',x)$ between 
$x'-\e$ and $x'+\e$ with $\e \to 0$, one obtains the second condition. Hence:
\begin{eqnarray}
A(x') K_{z}(x') &=& B(x') I_{z}(x'), \\
A(x')\, K_{z}'(x') - B(x')\, I_{z}'(x') &=& 1/x'. \nn
\end{eqnarray}
The two above conditions have a unique solution if the Wronskian is non-vanishing, which is indeed the case as 
\be
K_{z} I_{z}' - I_{z} K_{z}' = 1/x'.
\ee
With $A(x') = - I_{z}(x')$ and $B(x') = - K_{z}(x')$, all conditions on the Green's function are satisfied. 

Using this Green's function to
solve the vector field equations iteratively gives that
the $v$-component of the normalizable mode is
\begin{equation}
a_{(z)v} = a_{(-z)u}^{(0)} {k_{v} k^{2z} \over k_{u}} {\G(-z) \over 2^{2z}\, \G(z)}
\left ( 1 + \sigma^2 k_{\chi}^2 c_{vv} \right ) + a_{(-z)v}^{(0)}\, \sigma^{2} k_{\chi}^2 c_{uv}.
\end{equation}
The $u$ component of the normalizable mode is then
\begin{equation}
a_{(z)u} = a_{(-z)v}^{(0)}\, {k_{u}\, k^{2z} \over k_{v}}\, {\G(-z) \over 2^{2z} \G(z)}
\left (1 - \sigma^2 k_{\chi}^2 c_{uu} \right) + a_{(-z)u}^{(0)}\, \sigma^{2} k_{\chi}^2  c_{uv}.
\end{equation}
In these expressions,
\bea
c_{vv} &=& \frac{1-z}{2 z^{2} \Gamma(-z) \Gamma(z)}
\left [ \left( (1-z) - 2(3-2z)z^{2} \right) {\sqrt{\pi}\, \G(1-z)\, \G(2-2z) \over 4\, \G(5/2-z)} + S_{1} + S_{4} - 2(S_{3} + S_{6}) \right ];
\nn \\
c_{uv} &=& \frac{1-z}{2^{1+2z}
\G(1+z)^{2}} \left[ \frac{\sqrt{\pi}  \G(1-z) \G(2-2z)}{4 \G(5/2-z)} + S_{1} + S_{4} - 2(S_{3}+S_{6}) - 4z (S_{2}+S_{5}) \right ]; \nn \\
c_{uu} &=& \frac{1-z}{z^{2} \G(-z) \G(z)} \left [ \frac{1-z(4+z-2z^{2})}{2^{2z}(3+4z(-2+z))} \G(1-z)^{2}
+ S_{1} + S_{4} - 2 (S_{3}+S_{6})-4z(S_{2}+S_{5}) \right ]. \nn
\eea
The constants $c_{ab}$ relate to the numerical constants appearing in the two point functions in \eqref{point}.
The constants $S_{a}$ are given in terms of integral over Bessel functions as,
\begin{align*}
S_{1} & = \int_{0}^{\infty} dy\, y\, K_{z}(y)\, K_{z}(y) \int_{0}^{y} dy'\, (y')^{3-2z}\, I_{z}(y')\, K_{z}(y')\\
S_{2} & = \int_{0}^{\infty} dy\, y\, K_{z}(y)\, K_{z}(y) \int_{0}^{y} dy'\, (y')^{1-2z}\, I_{z}(y')\, K_{z}(y')\\
S_{3} & = \int_{0}^{\infty} dy\, y\, K_{z}(y)\, K_{z}(y) \int_{0}^{y} dy'\, (y')^{2-2z}\, I_{z}(y')\, K_{(z-1)}(y')\\
S_{4} & = \int_{0}^{\infty} dy\, y\, K_{z}(y)\, I_{z}(y) \int_{y}^{\infty} dy'\, (y')^{3-2z}\, K_{z}(y')\, K_{z}(y')\\
S_{5} & = \int_{0}^{\infty} dy\, y\, K_{z}(y)\, I_{z}(y) \int_{y}^{\infty} dy'\, (y')^{1-2z}\, K_{z}(y')\, K_{z}(y')\\
S_{6} & = \int_{0}^{\infty} dy\, y\, K_{z}(y)\, I_{z}(y) \int_{y}^{\infty} dy'\, (y')^{2-2z}\, K_{z}(y')\, K_{(z-1)}(y')
\end{align*}


\begin{thebibliography}{10}

\bibitem{Son:2008ye}
D.~T. Son, ``{Toward an AdS/cold atoms correspondence: a geometric realization
  of the Schroedinger symmetry},'' {\em Phys. Rev.} {\bf D78} (2008) 046003,
  \href{http://xxx.lanl.gov/abs/0804.3972}{{\tt 0804.3972}}.

\bibitem{Balasubramanian:2008dm}
K.~Balasubramanian and J.~McGreevy, ``{Gravity duals for non-relativistic
  CFTs},'' {\em Phys. Rev. Lett.} {\bf 101} (2008) 061601,
  \href{http://xxx.lanl.gov/abs/0804.4053}{{\tt 0804.4053}}.

\bibitem{Duval:1990hj}
C.~Duval, G.~W. Gibbons, and P.~Horvathy, ``{Celestial Mechanics, Conformal
  Structures, and Gravitational Waves},'' {\em Phys. Rev.} {\bf D43} (1991)
  3907--3922, \href{http://xxx.lanl.gov/abs/hep-th/0512188}{{\tt
  hep-th/0512188}}.

\bibitem{Duval:2008jg}
C.~Duval, M.~Hassaine, and P.~A. Horvathy, ``{The geometry of Schr\'odinger
  symmetry in gravity background/non-relativistic CFT},'' {\em Annals Phys.}
  {\bf 324} (2009) 1158--1167, \href{http://xxx.lanl.gov/abs/0809.3128}{{\tt
  0809.3128}}.

\bibitem{Cardy:1992tq}
J.~L. Cardy, ``{Critical exponents of the chiral Potts model from conformal
  field theory},'' {\em Nucl. Phys.} {\bf B389} (1993) 577--586,
  \href{http://xxx.lanl.gov/abs/hep-th/9210002}{{\tt hep-th/9210002}}.

\bibitem{Maldacena:2008wh}
J.~Maldacena, D.~Martelli, and Y.~Tachikawa, ``{Comments on string theory
  backgrounds with non- relativistic conformal symmetry},'' {\em JHEP} {\bf 10}
  (2008) 072, \href{http://xxx.lanl.gov/abs/0807.1100}{{\tt 0807.1100}}.

\bibitem{Hellerman:1997yu}
  S.~Hellerman and J.~Polchinski,
  ``Compactification in the lightlike limit,''
  Phys.\ Rev.\  D {\bf 59} (1999) 125002
  [arXiv:hep-th/9711037].

\bibitem{Balasubramanian:2010uw}
K.~Balasubramanian and J.~McGreevy, ``{The particle number in Galilean
  holography},'' \href{http://xxx.lanl.gov/abs/1007.2184}{{\tt 1007.2184}}.

\bibitem{Ostlund:1981zz}
S.~Ostlund, ``{Incommensurate and commensurate phases in asymmetric clock
  models},'' {\em Phys. Rev.} {\bf B24} (1981) 398--405.

\bibitem{Howes:1983mk}
S.~Howes, L.~P. Kadanoff, and M.~Den~Nijs, ``{Quantum model for
  commensurate-incommensurate transitions},'' {\em Nucl. Phys.} {\bf B215}
  (1983) 169--208.

\bibitem{Fateev:1985mm}
V.~A. Fateev and A.~B. Zamolodchikov, ``{Parafermionic Currents in the
  Two-Dimensional Conformal Quantum Field Theory and Selfdual Critical Points
  in Z(n) Invariant Statistical Systems},'' {\em Sov. Phys. JETP} {\bf 62}
  (1985) 215--225.

\bibitem{Albertini:1988ux}
G.~Albertini, B.~M. McCoy, J.~H.~H. Perk, and S.~Tang, ``{Excitation spectrum
  and order parameter for the integrable N state chiral Potts model},'' {\em
  Nucl. Phys.} {\bf B314} (1989) 741.

\bibitem{Baxter:1988me}
R.~J. Baxter, ``{The superintegrable chiral Potts model},'' {\em Phys. Lett.}
  {\bf A133} (1988) 185--189.

\bibitem{Baxter:1988xk}
R.~J. Baxter, ``{Free energy of the solvable chiral Potts model},'' {\em J.
  Statist. Phys.} {\bf 52} (1988) 639--667.

\bibitem{Donos:2010tu}
A.~Donos and J.~P. Gauntlett, ``{Lifshitz Solutions of D=10 and D=11
  supergravity},'' \href{http://xxx.lanl.gov/abs/1008.2062}{{\tt 1008.2062}}.

\bibitem{Guica:2010sw}
M.~Guica, K.~Skenderis, M.~Taylor, and B.~van Rees, ``{Holography for
  Schrodinger backgrounds},'' \href{http://xxx.lanl.gov/abs/1008.1991}{{\tt
  1008.1991}}.

\bibitem{Olmez:2005by}
S.~Olmez, O.~Sarioglu, and B.~Tekin, ``{Mass and angular momentum of
  asymptotically AdS or flat solutions in the topologically massive gravity},''
  {\em Class. Quant. Grav.} {\bf 22} (2005) 4355--4362,
  \href{http://xxx.lanl.gov/abs/gr-qc/0507003}{{\tt gr-qc/0507003}}.

\bibitem{Chow:2009km}
D.~D.~K. Chow, C.~N. Pope, and E.~Sezgin, ``{Classification of solutions in
  topologically massive gravity},'' {\em Class. Quant. Grav.} {\bf 27} (2010)
  105001, \href{http://xxx.lanl.gov/abs/0906.3559}{{\tt 0906.3559}}.

\bibitem{Anninos:2010pm}
D.~Anninos, G.~Compere, S.~de~Buyl, S.~Detournay, and M.~Guica, ``{The Curious
  Case of Null Warped Space},'' \href{http://xxx.lanl.gov/abs/1005.4072}{{\tt
  1005.4072}}.

\bibitem{Skenderis:2009nt}
K.~Skenderis, M.~Taylor, and B.~C. van Rees, ``{Topologically Massive Gravity
  and the AdS/CFT Correspondence},'' {\em JHEP} {\bf 09} (2009) 045,
  \href{http://xxx.lanl.gov/abs/0906.4926}{{\tt 0906.4926}}.

\bibitem{Skenderis:2009kd}
K.~Skenderis, M.~Taylor, and B.~C. van Rees, ``{AdS boundary conditions and the
  Topologically Massive Gravity/CFT correspondence},''
  \href{http://xxx.lanl.gov/abs/0909.5617}{{\tt 0909.5617}}.

\bibitem{Kachru:2008yh}
  S.~Kachru, X.~Liu and M.~Mulligan,
  ``Gravity Duals of Lifshitz-like Fixed Points,''
  Phys.\ Rev.\  D {\bf 78} (2008) 106005
  [arXiv:0808.1725 [hep-th]].




\bibitem{Gubser:2000nd}
  S.~S.~Gubser,
  ``Curvature singularities: The good, the bad, and the naked,''
  Adv.\ Theor.\ Math.\ Phys.\  {\bf 4} (2000) 679
  [arXiv:hep-th/0002160].


\bibitem{Leigh:2009eb}
R.~G. Leigh and N.~N. Hoang, ``{Real-Time Correlators and Non-Relativistic
  Holography},'' {\em JHEP} {\bf 11} (2009) 010,
  \href{http://xxx.lanl.gov/abs/0904.4270}{{\tt 0904.4270}}.

\bibitem{Barnes:2010ev}
E.~Barnes, D.~Vaman, and C.~Wu, ``{Holographic real-time non-relativistic
  correlators at zero and finite temperature},''
  \href{http://xxx.lanl.gov/abs/1007.1644}{{\tt 1007.1644}}.

\bibitem{Blau:2010fh}
M.~Blau, J.~Hartong, and B.~Rollier, ``{Geometry of Schroedinger Space-Times
  II: Particle and Field Probes of the Causal Structure},'' {\em JHEP} {\bf 07}
  (2010) 069, \href{http://xxx.lanl.gov/abs/1005.0760}{{\tt 1005.0760}}.

\bibitem{Hartnoll:2008rs}
S.~A. Hartnoll and K.~Yoshida, ``{Families of IIB duals for nonrelativistic
  CFTs},'' {\em JHEP} {\bf 12} (2008) 071,
  \href{http://xxx.lanl.gov/abs/0810.0298}{{\tt 0810.0298}}.

\bibitem{Gauntlett:2009zw}
J.~P. Gauntlett, S.~Kim, O.~Varela, and D.~Waldram, ``{Consistent
  supersymmetric Kaluza--Klein truncations with massive modes},'' {\em JHEP}
  {\bf 04} (2009) 102, \href{http://xxx.lanl.gov/abs/0901.0676}{{\tt
  0901.0676}}.

\bibitem{Donos:2009en}
A.~Donos and J.~P. Gauntlett, ``{Supersymmetric solutions for non-relativistic
  holography},'' {\em JHEP} {\bf 03} (2009) 138,
  \href{http://xxx.lanl.gov/abs/0901.0818}{{\tt 0901.0818}}.

\bibitem{Colgain:2009wm}
E.~O. Colgain and H.~Yavartanoo, ``{NR $CFT_3$ duals in M-theory},'' {\em JHEP}
  {\bf 09} (2009) 002, \href{http://xxx.lanl.gov/abs/0904.0588}{{\tt
  0904.0588}}.

\bibitem{Bobev:2009mw}
N.~Bobev, A.~Kundu, and K.~Pilch, ``{Supersymmetric IIB Solutions with
  Schr\'{o}dinger Symmetry},'' {\em JHEP} {\bf 07} (2009) 107,
  \href{http://xxx.lanl.gov/abs/0905.0673}{{\tt 0905.0673}}.

\bibitem{Donos:2009xc}
A.~Donos and J.~P. Gauntlett, ``{Solutions of type IIB and D=11 supergravity
  with Schrodinger(z) symmetry},'' {\em JHEP} {\bf 07} (2009) 042,
  \href{http://xxx.lanl.gov/abs/0905.1098}{{\tt 0905.1098}}.

\bibitem{Donos:2009zf}
A.~Donos and J.~P. Gauntlett, ``{Schrodinger invariant solutions of type IIB
  with enhanced supersymmetry},'' {\em JHEP} {\bf 10} (2009) 073,
  \href{http://xxx.lanl.gov/abs/0907.1761}{{\tt 0907.1761}}.

\bibitem{Singh1}
H.~Singh,
  ``Galilean anti-de-Sitter spacetime in Romans theory,''
  Phys.\ Lett.\  {\bf B682}, 225-228 (2009).
  [arXiv:0909.1692 [hep-th]].

\bibitem{Singh2}
H.~Singh,
  ``String dual of a Bose-Einstein condensate,''
  [arXiv:1009.0651 [hep-th]].


\bibitem{Kim:1985ez}
H.~J. Kim, L.~J. Romans, and P.~van Nieuwenhuizen, ``{The Mass Spectrum of
  Chiral N=2 D=10 Supergravity on S**5},'' {\em Phys. Rev.} {\bf D32} (1985)
  389.

\bibitem{Ceresole:1999zs}
A.~Ceresole, G.~Dall'Agata, R.~D'Auria, and S.~Ferrara, ``{Spectrum of type IIB
  supergravity on AdS(5) x T(11): Predictions on N = 1 SCFT's},'' {\em Phys.
  Rev.} {\bf D61} (2000) 066001,
  \href{http://xxx.lanl.gov/abs/hep-th/9905226}{{\tt hep-th/9905226}}.

\bibitem{Ceresole:1999ht}
A.~Ceresole, G.~Dall'Agata, and R.~D'Auria, ``{KK spectroscopy of type IIB
  supergravity on AdS(5) x T(11)},'' {\em JHEP} {\bf 11} (1999) 009,
  \href{http://xxx.lanl.gov/abs/hep-th/9907216}{{\tt hep-th/9907216}}.

\bibitem{Balasubramanian:2010uk}
  K.~Balasubramanian and K.~Narayan,
  ``Lifshitz spacetimes from AdS null and cosmological solutions,''
  JHEP {\bf 1008} (2010) 014
  [arXiv:1005.3291 [hep-th]].

\bibitem{Narayan:2011az}
  K.~Narayan,
  ``Lifshitz-like systems and AdS null deformations,''
  arXiv:1103.1279 [hep-th].

\bibitem{Taylor:2008tg}
M.~Taylor, ``{Non-relativistic holography},''
  \href{http://xxx.lanl.gov/abs/0812.0530}{{\tt 0812.0530}}.

\bibitem{Kanitscheider:2008kd}
I.~Kanitscheider, K.~Skenderis, and M.~Taylor, ``{Precision holography for
  non-conformal branes},'' {\em JHEP} {\bf 09} (2008) 094,
  \href{http://xxx.lanl.gov/abs/0807.3324}{{\tt 0807.3324}}.

\bibitem{Kanitscheider:2009as}
I.~Kanitscheider and K.~Skenderis, ``{Universal hydrodynamics of non-conformal
  branes},'' {\em JHEP} {\bf 04} (2009) 062,
  \href{http://xxx.lanl.gov/abs/0901.1487}{{\tt 0901.1487}}.

\bibitem{Dijkgraaf:1996iy}
R.~Dijkgraaf, ``{Chiral deformations of conformal field theories},'' {\em Nucl.
  Phys.} {\bf B493} (1997) 588--612,
  \href{http://xxx.lanl.gov/abs/hep-th/9609022}{{\tt hep-th/9609022}}.

\bibitem{Freedman:1991tk}
D.~Z. Freedman, K.~Johnson, and J.~I. Latorre, ``{Differential regularization
  and renormalization: A New method of calculation in quantum field theory},''
  {\em Nucl. Phys.} {\bf B371} (1992) 353--414.

\bibitem{Freedman:1992gr}
D.~Z. Freedman, K.~Johnson, R.~Munoz-Tapia, and X.~Vilasis-Cardona, ``{A Cutoff
  procedure and counterterms for differential renormalization},'' {\em Nucl.
  Phys.} {\bf B395} (1993) 454--496,
  \href{http://xxx.lanl.gov/abs/hep-th/9206028}{{\tt hep-th/9206028}}.

\bibitem{Skenderis:2002wp}
K.~Skenderis, ``Lecture notes on holographic renormalization,'' {\em Class.
  Quant. Grav.} {\bf 19} (2002) 5849--5876,
  \href{http://xxx.lanl.gov/abs/hep-th/0209067}{{\tt hep-th/0209067}}.

\bibitem{Papadimitriou:2004ap}
I.~Papadimitriou and K.~Skenderis, ``{AdS / CFT correspondence and geometry},''
  {\em Proceedings Strasbourg 2003, AdS/CFT correspondence, ed. O. Biquard} 73,
  \href{http://xxx.lanl.gov/abs/hep-th/0404176}{{\tt hep-th/0404176}}.

\bibitem{Papadimitriou:2005ii}
I.~Papadimitriou and K.~Skenderis, ``{Thermodynamics of asymptotically locally
  AdS spacetimes},'' {\em JHEP} {\bf 08} (2005) 004,
  \href{http://xxx.lanl.gov/abs/hep-th/0505190}{{\tt hep-th/0505190}}.

\bibitem{deHaro:2000xn}
S.~de~Haro, S.~N. Solodukhin, and K.~Skenderis, ``{Holographic reconstruction
  of spacetime and renormalization in the AdS/CFT correspondence},'' {\em
  Commun. Math. Phys.} {\bf 217} (2001) 595--622,
  \href{http://xxx.lanl.gov/abs/hep-th/0002230}{{\tt hep-th/0002230}}.

\bibitem{Henningson:1998gx}
M.~Henningson and K.~Skenderis, ``{The holographic Weyl anomaly},'' {\em JHEP}
  {\bf 07} (1998) 023, \href{http://xxx.lanl.gov/abs/hep-th/9806087}{{\tt
  hep-th/9806087}}.

\bibitem{Brown:1994py}
J.~D. Brown and K.~V. Kuchar, ``{Dust as a standard of space and time in
  canonical quantum gravity},'' {\em Phys. Rev.} {\bf D51} (1995) 5600--5629,
  \href{http://xxx.lanl.gov/abs/gr-qc/9409001}{{\tt gr-qc/9409001}}.

\bibitem{Brown:1994ix}
J.~D. Brown, ``{On variational principles for gravitating perfect fluids},''
  \href{http://xxx.lanl.gov/abs/gr-qc/9407008}{{\tt gr-qc/9407008}}.

\bibitem{Compere:2010uk}
G.~Compere, W.~Song, and A.~Virmani, ``{Microscopics of Extremal Kerr from
  Spinning M5 Branes},'' \href{http://xxx.lanl.gov/abs/1010.0685}{{\tt
  1010.0685}}.

\bibitem{Guica:2010ej}
M.~Guica and A.~Strominger, ``{Microscopic Realization of the Kerr/CFT
  Correspondence},'' \href{http://xxx.lanl.gov/abs/1009.5039}{{\tt 1009.5039}}.

\end{thebibliography}

\end{document}